\documentclass[]{article}
\makeatletter\if@twocolumn\PassOptionsToPackage{switch}{lineno}\else\fi\makeatother

\usepackage{amsfonts,amssymb,amsbsy,latexsym,amsmath,tabulary,graphicx,times,caption,fancyhdr}
\usepackage[utf8]{inputenc}
\usepackage{url,multirow,morefloats,floatflt,cancel,tfrupee}
\usepackage{setspace}

\usepackage[multiple]{footmisc}
\usepackage{footnotebackref}

\usepackage{hyperref}
\hypersetup{
    colorlinks=true,
    linkcolor=blue,
    filecolor=magenta, 
    citecolor=blue,
    urlcolor=blue,
    pdftitle={Overleaf Example},
    pdfpagemode=FullScreen,
    }
\usepackage{algorithm,algpseudocode}

\makeatletter

\AtBeginDocument{\@ifpackageloaded{textcomp}{}{\usepackage{textcomp}}}
\makeatother
\usepackage{colortbl}
\usepackage{xcolor}
\usepackage{pifont}
\usepackage[nointegrals]{wasysym}
\usepackage{lscape}

\makeatletter

\def\mcWidth#1{\csname TY@F#1\endcsname+\tabcolsep}

\def\cAlignHack{\rightskip\@flushglue\leftskip\@flushglue\parindent\z@\parfillskip\z@skip}
\def\rAlignHack{\rightskip\z@skip\leftskip\@flushglue \parindent\z@\parfillskip\z@skip}

\@ifundefined{etal}{}{}

\usepackage{ifxetex}
\ifxetex\else\if@twocolumn\@ifpackageloaded{stfloats}{}{\usepackage{dblfloatfix}}\fi\fi

\AtBeginDocument{
\expandafter\ifx\csname eqalign\endcsname\relax
\def\eqalign#1{\null\vcenter{\def\\{\cr}\openup\jot\m@th
  \ialign{\strut$\displaystyle{##}$\hfil&$\displaystyle{{}##}$\hfil
      \crcr#1\crcr}}\,}
\fi
}

\AtBeginDocument{%
  \@ifpackageloaded{endfloat}%
   {\renewcommand\efloat@iwrite[1]{\immediate\expandafter\protected@write\csname efloat@post#1\endcsname{}}}{\newif\ifefloat@tables}%
}%

\def\BreakURLText#1{\@tfor\brk@tempa:=#1\do{\brk@tempa\hskip0pt}}
\let\lt=<
\let\gt=>
\def\processVert{\ifmmode|\else\textbar\fi}

\@ifundefined{subparagraph}{
\def\subparagraph{\@startsection{paragraph}{5}{2\parindent}{0ex plus 0.1ex minus 0.1ex}%
{0ex}{\normalfont\small\itshape}}%
}{}

\newcommand\role[1]{\unskip}
\newcommand\aucollab[1]{\unskip}
  
\@ifundefined{tsGraphicsScaleX}{\gdef\tsGraphicsScaleX{1}}{}
\@ifundefined{tsGraphicsScaleY}{\gdef\tsGraphicsScaleY{.9}}{}
\def\checkGraphicsWidth{\ifdim\Gin@nat@width>\linewidth
	\tsGraphicsScaleX\linewidth\else\Gin@nat@width\fi}

\def\checkGraphicsHeight{\ifdim\Gin@nat@height>.9\textheight
	\tsGraphicsScaleY\textheight\else\Gin@nat@height\fi}

\def\fixFloatSize#1{}
\let\ts@includegraphics\includegraphics

\def\inlinegraphic[#1]#2{{\edef\@tempa{#1}\edef\baseline@shift{\ifx\@tempa\@empty0\else#1\fi}\edef\tempZ{\the\numexpr(\numexpr(\baseline@shift*\f@size/100))}\protect\raisebox{\tempZ pt}{\ts@includegraphics{#2}}}}

\AtBeginDocument{\def\includegraphics{\@ifnextchar[{\ts@includegraphics}{\ts@includegraphics[width=\checkGraphicsWidth,height=\checkGraphicsHeight,keepaspectratio]}}}

\DeclareMathAlphabet{\mathpzc}{OT1}{pzc}{m}{it}

\def\URL#1#2{\@ifundefined{href}{#2}{\href{#1}{#2}}}

\def\UrlOrds{\do\*\do\-\do\~\do\'\do\"\do\-}%
\g@addto@macro{\UrlBreaks}{\UrlOrds}

\edef\fntEncoding{\f@encoding}

\makeatother

\newif\ifmultipleabstract\multipleabstractfalse%
\makeatletter

\def\wileyIndent{1pt}
\usepackage[paperheight=10in,paperwidth=6.5in,margin=1.5cm,headsep=.5cm,top=2.5cm,headheight=1cm]{geometry}

\renewenvironment{abstract}
{\vspace*{-1pc}\trivlist\item[]\leftskip\wileyIndent\hrulefill\par\vskip4pt\noindent\textbf{\abstractname}\mbox{\null}\\}{\par\noindent\hrulefill\endtrivlist}

\def\author#1{\gdef\@author{\hskip-\dimexpr(\tabcolsep)\hskip\wileyIndent\parbox{\dimexpr\textwidth-\wileyIndent}{\centering\bfseries#1}}}

\def\title#1{\linespread{1}\gdef\@title{\centering\bfseries\ifx\@articleType\@empty\else\@articleType\\\fi#1}}

\let\@articleType\@empty \def\articletype#1{\gdef\@articleType{{\normalfont\itshape#1}}}

\AtBeginDocument{\fancypagestyle{headings}{\fancyhf{}\fancyhead[C]{\RunningHead}\fancyhead[R]{\thepage}}\pagestyle{headings}}

 \def\audegree#1{}

\date{}

\usepackage{multirow}

\makeatother

\usepackage[T1]{fontenc}
\makeatother
\usepackage[authoryear,round]{natbib}

\usepackage{float}

\DeclareMathOperator{\argmin}{argmin}

\newtheorem{algo}{Algorithm}[section]

\usepackage{subfig}
\usepackage{tablefootnote}

\let\oldFootnote\footnote
\newcommand\nextToken\relax

\renewcommand\footnote[1]{%
    \oldFootnote{#1}\futurelet\nextToken\isFootnote}

\newcommand\isFootnote{%
    \ifx\footnote\nextToken\textsuperscript{,}\fi}

\usepackage{titlesec}
\titlelabel{\thetitle.\quad}

\begin{document}

\title{\textbf{Fast Estimation of Bayesian State Space Models Using Amortized Simulation-Based Inference }\footnote{The views expressed in the paper are solely those of the authors and do not necessarily represent the official position of the Bank of Russia. The Bank of Russia is not responsible for the contents of the paper.} \footnote{We are grateful to Dmitry Gornostaev, Sergey Ivaschenko, Petr Milyutin, Denis Shibitov and participants of the 15th International Conference on Computational and Financial Econometrics (CEF 2021), XXIII Yasin (April) International Academic Conference on Economic and Social Development and the 2nd International Conference on Econometrics and Business Analytics (iCEBA) for their helpful comments and suggestions.}}
\author{Ramis~Khabibullin\textsuperscript{1}\space and Sergei~Seleznev\textsuperscript{2}~\\[-3pt]\normalsize\normalfont  \itshape ~\\
\textsuperscript{1}{Bank of Russia}~\\
\textsuperscript{2}{Bank of Russia}}

\def\RunningHead{}\def\RunningAuthor{Khabibullin and Seleznev}

\maketitle

\begin{abstract}
This paper presents a fast algorithm for estimating hidden states of Bayesian state space models. The algorithm is a variation of amortized simulation-based inference algorithms, where numerous artificial datasets are generated at the first stage, and then a flexible model is trained to predict the variables of interest. In contrast to those proposed earlier, the procedure described in this paper makes it possible to train estimators for hidden states by concentrating only on certain characteristics of the marginal posterior distributions and introducing inductive bias.

\noindent Illustrations using the examples of stochastic volatility model, nonlinear dynamic stochastic general equilibrium model and seasonal adjustment procedure with breaks in seasonality show that the algorithm has sufficient accuracy for practical use. Moreover, after pretraining, which takes several hours, finding the posterior distribution for any dataset takes from hundredths to tenths of a second.

\smallskip\noindent\textbf{JEL-classification: C11, C15, C32, C45.}\def\keywordstitle{Keywords}

\end{abstract}
    
\section{Introduction}
Bayesian state space models are widely used in applied macroeconomics. They are so widespread due to the fact that many macroeconomic and econometric models can be written in the form of state space models for subsequent estimation on real data. For example, various kinds of filters and semi-structural filters (see \cite{hodrick1997postwar}, \cite{laubach2003}), models with stochastic volatility (see \cite{kim1998stochastic}, \cite{justiniano2008time}, \cite{carriero2019large}), time-varying models (see \cite{10.2307/1912559}, \cite{primiceri2005time}, \cite{korobilis}), mixed frequency models (see \cite{chiu2012estimating}, \cite{schorfheide2015, schorfheide2021real}), dynamic factor models (see \cite{otrok1998bayesian}, \cite{stock2011dynamic}), dynamic stochastic general equilibrium models (see \cite{smets2003estimated,smets2007shocks}, \cite{FERNANDEZVILLAVERDE2016527}) and agent-based models (see \cite{lux2018estimation}, \cite{gatti2020rising}). Bayesian parameter estimation makes it possible to mitigate the lack of long time series\footnote{In this paper, we focus on time series models, but the proposed algorithm can easily be transferred to hidden space models of other types with minor modifications in the architecture.}.

Despite their flexibility, in practice, estimating Bayesian state space models is a rather difficult task. Sampling algorithms are based on an iterative sampling scheme for model parameters and states and rely on steps such as Gibbs sampling (see \cite{casella1992explaining}), Metropolis-Hastings (see \cite{chib1995understanding}), Hamiltonian Monte Carlo (see \cite{neal2012bayesian}) or sequential Monte Carlo (see \cite{del2006sequential}). Even in cases where the model is linear and Gaussian or discrete with respect to states, sampling can take from tens of minutes to hours. In systems of a more general form, researchers use particle filters (see  \cite{andrieu2010particle}, \cite{chopin2013smc2}), as a result of which estimation time only increases. Sampling algorithms are exact in the sense that they converge to the posterior distribution as the number of iterations tends to infinity, although the number of iterations required for an estimate close to the posterior distribution can be large. An alternative to them are optimization algorithms. The most common of them in the context of state space models is the variational Bayes algorithm\footnote{See Chapters 3 and 5 of \cite{beal2003variational} and \cite{gunawan2021variational} for examples of using the Bayesian variational algorithm in estimating of state space models.} (see \cite{wainwright}, \cite{hoffman2013stochastic}). The variational Bayes algorithm relies on minimizing the Kullback-Leibler (or any other) divergence between the approximation and the true posterior distribution. It is often faster than sampling algorithms, but its running time is also large, especially in cases where the optimization steps cannot be written in analytical form.

In this paper, we propose a fast algorithm for estimating Bayesian state space models that is based on the principles of simulation-based inference (see \cite{cranmer2020frontier}). The speed of the algorithm is achieved by amortizing the task of constructing the posterior distribution of states, that is, by pretraining a model (a neural network, in our case) that predicts its posterior distribution from the data. In doing so, we focus only on states and do this for two reasons. First, in many problems it is the states, not the model parameters, that are of particular interest. For example, in the detrending problem (see \cite{orphanides2002unreliability}), the trend and cycle components, which are determined by hidden states, are of main interest. Second, simulation-based inference parameter estimation has been investigated in many other papers (see Appendix A of \cite{lueckmann2021benchmarking}) and can be easily combined with the approach discussed in this paper. The problem of constructing posterior distribution of hidden states is much more complicated due to its dimensionality and has not been studied much in the literature.

Section \ref{sec::2est_proc} describes the algorithm for estimating the posterior distribution of the model. Section \ref{sec::3appl} is devoted to the study of application, practical characteristics and comparison of the algorithm with commonly used alternatives. Section \ref{sec::4rela_wrk} describes related work. Section \ref{sec::5disc} discusses issues that have not been included in the paper but are important in the context of the proposed algorithm. The conclusion is presented in Section \ref{sec::6concl}.

\section{Estimation procedure} \label{sec::2est_proc}
\subsection{Amortized simulation-based inference}\label{subsec::2_1amort_inf}
Our methodology for estimating state space models is based on the idea of estimating Bayesian parameters proposed by \cite{beaumont2002approximate}, \cite{blum2010non} and developed by \cite{papamakarios2016fast}. The essence of the methodology is to simulate the joint distribution of parameters and data, and then predict the parameters conditional on data.

Formally, during the first step, a dataset is simulated from a model with a prior distribution $p(\theta)$ and a likelihood function $p(y | \theta)$. The $i$-th point consists of parameters $\theta_i \sim p(\theta)$ and observed variables $y_i \sim p(y | \theta_i)$. In the second step, an estimator is fitted to predict the distribution $p(\theta | y)$ or its characteristics, for example, by minimizing the cross-entropy between the simulated data and some parametric family of distributions $q_{\varphi} (\theta|y) p(y)$:
\begin{equation}
    \varphi^{*} = \argmin_{\varphi} \left[
    - \sum_{i = 1}^{N}
    \biggl( log\,q_{\varphi} (\theta_i | y_i) + log\,p(y_i)\biggr)
    \right]
\end{equation}

\noindent where $\varphi$ is the vector of parameters of distribution $q_{\varphi} (\theta | y)$, $N$ is the number of  simulations. We will omit $log\,p(y_i)$ term due to its independence from $\varphi$ in the following. 

The estimated distribution will tend to the posterior for any $y$ with an infinite number of simulations and a sufficiently flexible parametric family $q_{\varphi}$ (see Appendix \ref{app:A1} for an informal proof). This means that once estimated, the conditional distribution $q_{\varphi^{*}} (\theta | y) $ can be used for any data, does not require re-estimation of the model and can be calculated almost instantly.

As can be seen, this method for estimating posterior distributions (hereinafter, we will call it NPE following \cite{papamakarios2016fast}) does not require knowledge of $p(\theta)$ and $p(y|\theta)$ in an explicit form but relies only on the ability of simulating data, which is natural for most models. So, it falls into the category of\textit{ simulation-based inference} (SBI) methods (or \textit{likelihood-free inference}).

\subsection{From Bayesian models to state space models}\label{subsec::2_2FromBayes2SS}
The parameters of state space models can be estimated using the procedure presented in Subsection \ref{subsec::2_1amort_inf}, however the goal of this paper is to estimate the hidden states. It is easy to see that if we replace the parameters with hidden states in the loss function, then the procedure described above remains valid (see Appendix \ref{app:A2}). Thus, in general, the algorithm for finding the posterior distribution can be written as shown below:

\begin{algo}{\textbf{Simulation based state space model inference}}\label{algo::sbl_ss}
\vspace{0.5ex}\hrule\leavevmode \\
\normalfont  
{For} $i = 1, \ldots, N$:
\begin{enumerate}
    \item	Simulate states and observable data:
    \begin{enumerate}
        \item Draw model parameters from the prior:\\
        \begin{minipage}{.5\textwidth}
        $$ \theta_i \sim p(\theta) $$
        \end{minipage}
        \item Draw states from the conditional state distribution:\\
        \begin{minipage}{.5\textwidth}
        $$ \,\,\,s_i \sim p(s | \theta_i) $$
        \end{minipage}
        \item Draw data from the conditional data distribution:\\
        \begin{minipage}{.5\textwidth}
        $$ \,\,\,\,\,\,\,\,\,\,\,y_i \sim p(y | s_i, \theta_i) $$
        \end{minipage}
    \end{enumerate}
    \item Find the parameters of the posterior approximation of hidden states:
    \begin{equation}
    \varphi^{*} = \argmin_{\varphi} \left[
    - \sum_{i = 1}^{N}
     log \, q_{\varphi} (s_i | y_i) 
     \right]
\end{equation}
\end{enumerate}
\end{algo}
{\footnotesize
}
\,\,\,\,\,\, Despite the simplicity of Algorithm \ref{algo::sbl_ss}, there are several practical difficulties when moving from finding posterior distribution for the parameters to finding the distribution of states. First, it is a large dimension of the hidden space. Despite the presence of various kinds of flow transformations (see \cite{rezende2015variational}), which are often used in SBI and in many cases recover joint distributions adequately, their application for problems of this size, complicated by amortization, is costly to compute and is associated with optimization challenges. Second, it is the large dimensionality of the data. It is nearly impossible to find summary statistics that reduce the dimensionality of the data for hidden states, in contrast to the problem of parameter estimation, where this is common practice (see, for example, SIR (T.9) and Lotka-Voltera (T.10) models in  \cite{lueckmann2021benchmarking}). To overcome these problems and simplify the task of training the model, we avoid modeling the dependencies between variables focusing on characteristics of marginal distributions and introduce an inductive bias for $q_{\varphi} (s| y)$.

\subsection{Marginal distribution loss}\label{subsec::2_3marglik}
Moving from the joint to the marginal distributions is equivalent to subdividing the task into a set of one-dimensional tasks. In such case, the log-probability for the parametric family $q_{\varphi} (s | y)$ is written as:
\begin{equation}
    log\,q_{\varphi} = \sum_{t=1}^{T} \sum_{k=1}^{K} log \, q_{\varphi_{t,k}} \left(s^{t, k} | y \right) 
\end{equation}
where $t$ and $k$ represent the time period and the state index in state vector. Note that for each state, the vector of parameters $\varphi_{t,k}$ is generally its own and $\varphi$ consists of a set of these vectors.

We use the normal distribution with mean $m_{\varphi_{t,k}} (y)$ and standard deviation $\sigma_{\varphi_{t,k}} (y)$  for $q_{\varphi_{t,k}} (s^{t,k} | y)$. It is easy to show that for sufficiently flexible $m_{\varphi_{t,k} } (y)$ and $\sigma_{\varphi_{t,k}} (y)$ such an approximation exactly recovers the mean and standard deviation of the true posterior distribution (see Appendix \ref{app:A3}).

Moreover, the choice of normal distribution and cross-entropy as a loss function can be relaxed. Therefore, a mixture of normals or a small-scale flow-based model can replace normal distribution for marginal densities and any M-estimator (see Chapter 5 in \cite{van2000asymptotic}), such as quantile regression loss, can be used instead of cross-entropy.

\subsection{Estimator architecture}\label{subsec::2_4est_arch}
A natural architecture that is similar to filtering and smoothing in state space models is a Bidirectional Recurrent Neural Network (Bidirectional RNN). By sharing parameters for models with a «near» stationary data generation process\footnote{We use the term «near» stationary data generation process to emphasize the possibility of using different state distributions for the initial time period.}\footnote{One can always connect several networks, if there is a shift in the data generation process.}, this structure allows to move away from estimating the neural network for each state separately and calculate the loss function for all states in a single pass over the data. 

To make the network architecture more flexible, we use data convolution of various lengths as RNN input in addition to the raw data, and transform the output of RNN by applying a linear transformation or a fully connected neural network.

\section{Applications}\label{sec::3appl}
To illustrate the properties of the proposed method, we estimate three models: stochastic volatility model, non-linear DSGE model and seasonal adjustment model with structural breaks in seasonality.

\subsection{Stochastic volatility model}\label{subsec::3_1sv}

The stochastic volatility model (see \cite{kim1998stochastic}) is a classic example for testing various states estimation algorithms in state space models (see \cite{tan2020conditionally}). The model specification is exactly the same as \cite{tan2020conditionally} and presented in Appendix \ref{app:B1}. To simplify the learning process, the logarithm of absolute values of the observed data is fed into the neural network as an input (details of the network architecture and the training algorithm are presented in Appendix \ref{app:B2}). The model is trained on 20,000,000 generated series with a length of 800 to 1,200 observations.

\begin{figure}[H]
    \centering
    \includegraphics[width=\linewidth]{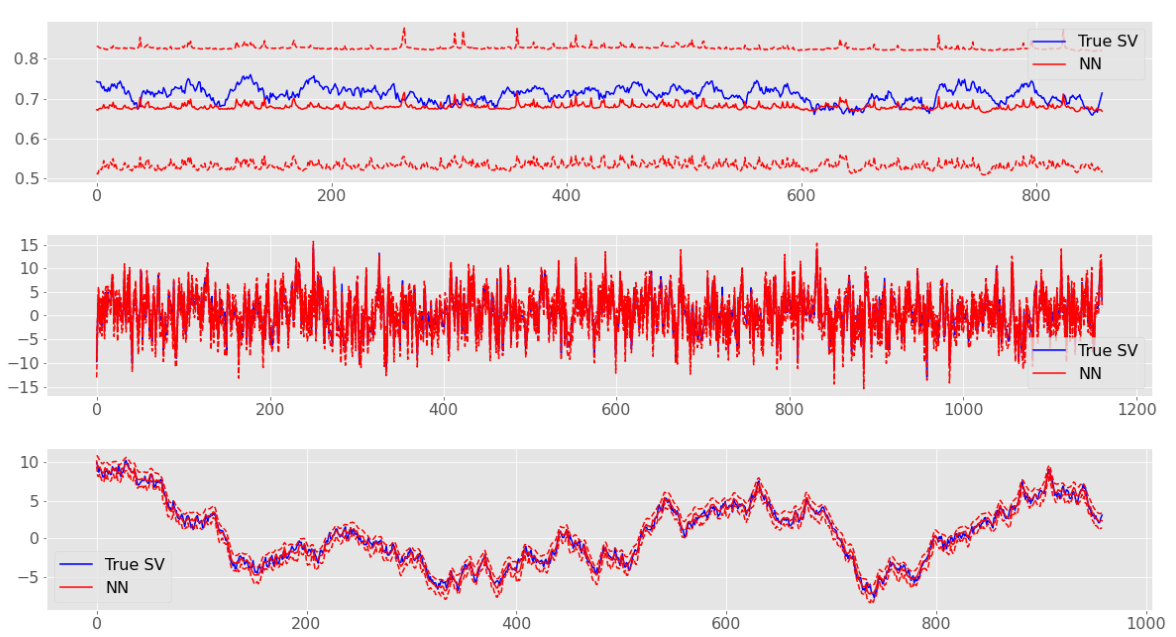}
    \caption{True values of volatility logarithms and approximation of posterior distribution on artificial data (mean $\pm$ 2 std)}
    \label{fig::art0}
\end{figure}

First, we demonstrate the quality of the algorithm on the data sampled from the data generation process. Figure \ref{fig::art0} shows examples of the true values of volatility logarithms and mean of posterior distribution approximation ($\pm$ 2 standard deviations). The figure demonstrates that the true values are quite well estimated by the neural network. As a benchmark for future studies, we also present the negative log-likelihood (hereinafter, NLL) and mean squared error (hereinafter, MSE) on a randomly generated 1,000,000 runs in Table \ref{tab::time_NLL}.

\begin{table}[H]
\centering
    \begin{tabular}{c| c c}
    & {{NLL} }& {{MSE}} \\
    \hline
   {{SV}} & $(8.17\pm0.21)\times10^{-2}$ &	$(2.92\pm0.02)\times10^{-1}$ \\
   {{SV-DSGE}} &	$(-1.64\pm0.00)\times10^{-1}$&	$(4.77\pm0.00)\times10^{-2}$\\
   {{SA}} &	$-3.00\pm0.01$	& $(1.27\pm0.02)\times10^{-3}$ \\
   \hline
    \end{tabular}
    \caption{NLL and MSE for various applications (mean $\pm$ 2 std)}\label{tab::time_NLL}
\end{table}

Unfortunately, calculation of the accuracy metrics for other algorithms, such as MCMC or stochastic variational Bayes, is difficult to compute (see the discussion on quality metrics for SBI in \cite{lueckmann2021benchmarking}). Our focus therefore lies on the comparison of the results for NYSE and GBPUSD datasets similarly to \cite{tan2020conditionally}\footnote{Calculating metrics, such as C2ST, are not informative for the joint distribution. Information about correlations is important for classification, but the algorithm used assumes the diagonal covariance matrix. Calculation of C2ST for marginal distributions requires training the number of classifiers equal to the number of hidden states. So, we use only visual analysis.}. The results are compared with the adaptive MCMC algorithm based on the mixture of normals for chi-square distribution approximation, which was proposed by \cite{kim1998stochastic}, and the stochastic variational Gaussian approximation (hereinafter, VB) with a sparse precision matrix (both algorithms are given in Appendix \ref{app:B3}). As can be seen in Figure \ref{fig::sv0}, although the neural network estimates are a bit noisy, they are close to the MCMC algorithm, which serves as the gold standard, as well as the variational Bayes algorithm, which is one of the fastest and most accurate approximations. It should be noted that the NYSE dataset is almost twice the maximum size of the simulation, and the neural network has never seen data of that length. Nevertheless, the trained model copes with this task.

\begin{figure}[H]
    \centering
    \includegraphics[width=\linewidth]{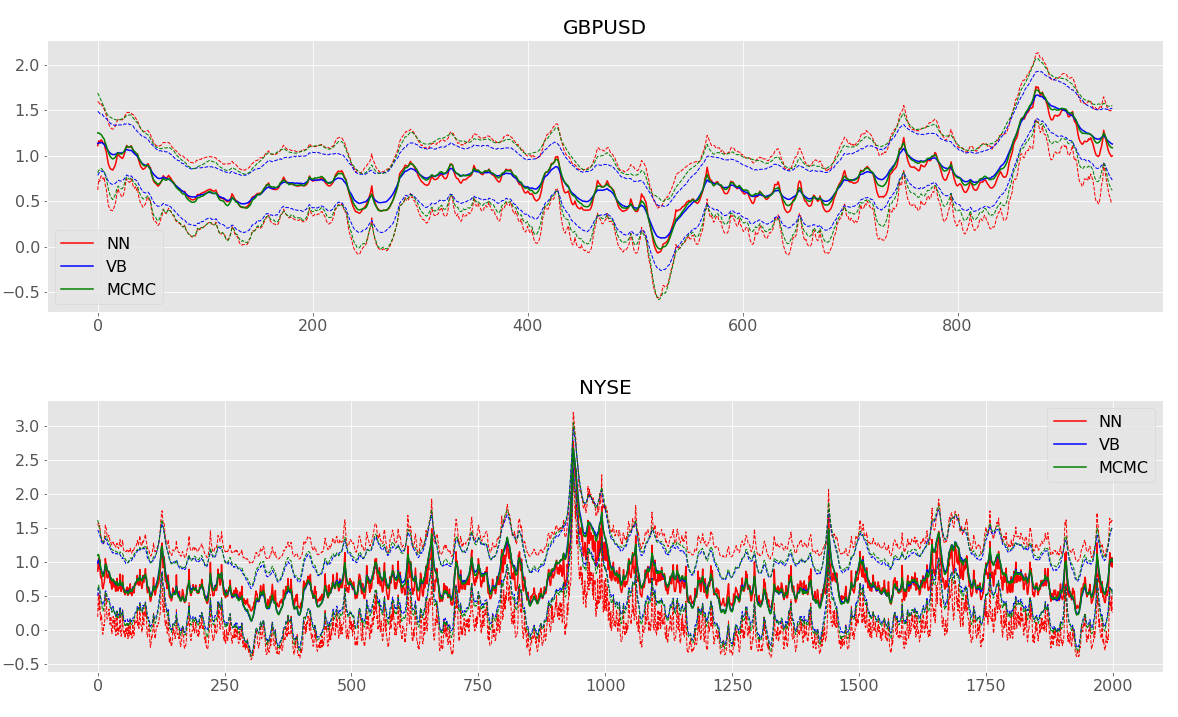}
    \caption{Comparison of NPE with MCMC and VB on real data for stochastic volatility model (mean $\pm$ 2 std)}
    \label{fig::sv0}
\end{figure}

\subsection{Stochastic volatility DSGE model}\label{subsec::3_2dsge}
The stochastic volatility model, while contains many hidden states, is a univariate model, both in terms of data and output\footnote{The latter could potentially be an advantage, though, as information from different sources can help in training neural network parameters. Moreover, it is possible to train a univariate model for each dimension of state vector, which reduces the problem to the previous one in terms of output.}. The DSGE model was chosen to test the amortized SBI algorithm in multivariate context. This class of models is widely used by macroeconomists, both for practical purposes (see \cite{linde2016challenges}) and in academic research (see \cite{walsh2010}). Although solving non-linear DSGE models is beyond the scope of this paper, we seek to demonstrate that the proposed algorithm works well for models where filtering and likelihood estimation cannot be executed using the Kalman filter as in the case of linear models\footnote{Log-linearized versions of these models are often used to overcome the computational difficulties with solving and estimating non-linear models.}. A simplified DSGE model\footnote{We exclude the inflation target shock from the model and inflation expectations from the observed variables to avoid the issues of missing variables and their impact on the result. A number of additional experiments have shown that using trained values for neural network inputs in place of missing variables (see \cite{lueckmann2017flexible}) and introducing additional dummy variables to the RNN input can cope with this task. However, we will focus here on a simpler version of the model to separate the effect of multiple observed variables from the effect of missing data. Estimates of DSGE models using SBI will be the subject of a separate paper where we will also touch on this issue.}  with stochastic volatility from \cite{diebold2017real} was chosen for these reasons. This model can be solved using standard algorithms for linearized models (see \cite{blanchard1980solution}, \cite{anderson1985linear}, \cite{klein2000using}, \cite{sims2002solving}). Nonlinearity is introduced after the solution step as the time-varying volatility of model shocks. The neural network is estimated on 50,000,000 generated datasets with a length of 180 to 200 points and compared with the adaptive MCMC algorithm (model description, neural network architecture and MCMC implementation are described in  \ref{app::C}).

\begin{figure}\centering
\subfloat{\label{a}\includegraphics[width=.45\linewidth]{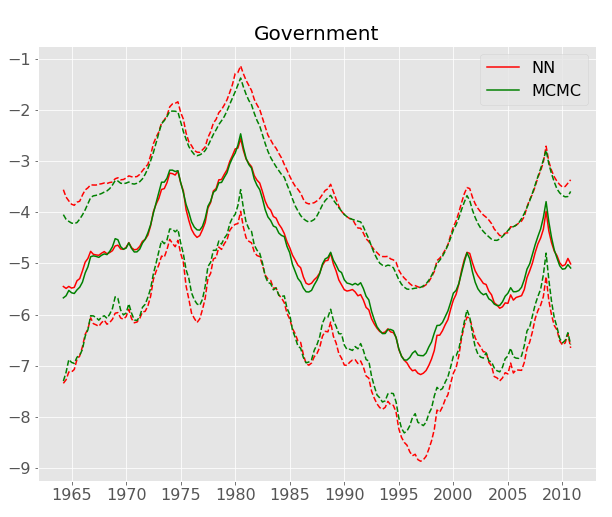}}\hfill
\subfloat{\label{b}\includegraphics[width=.45\linewidth]{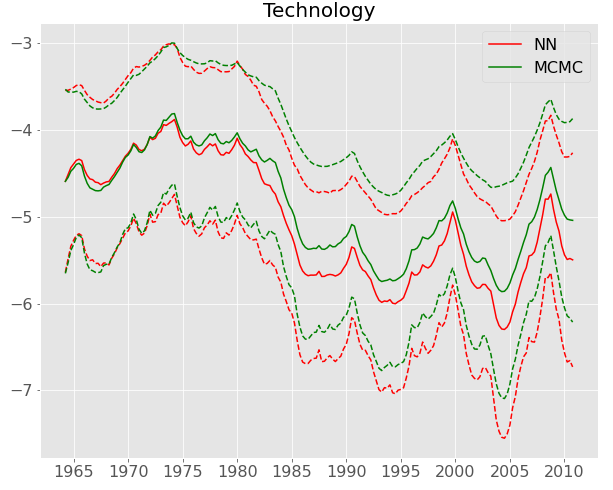}}\par 
\subfloat{\label{c}\includegraphics[width=.45\linewidth]{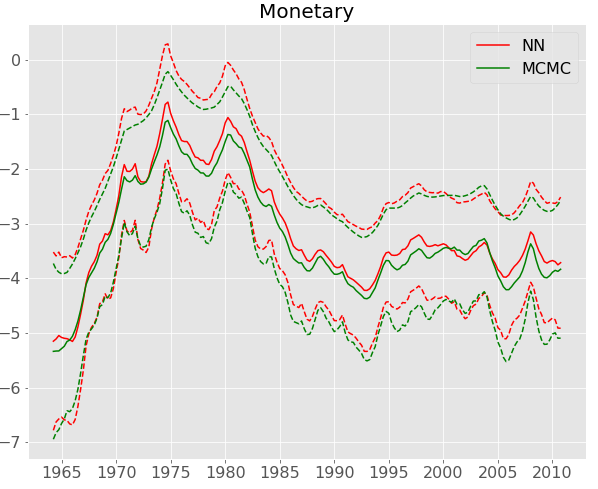}}
\caption{Comparison of NPE with MCMC on US data for DSGE model with stochastic volatility, stochastic volatility (mean $\pm$ 2 std)}
\label{fig::dsge_sv1}
\end{figure}

To illustrate the properties of NPE, we concentrated on the estimation of stochastic volatilities and present graphs and metrics for these states. However, unobservable shocks (more precisely, the logarithms of their absolute values) were also used in the estimation as a hint on the intermediate outputs of the neural network. Figure \ref{fig::dsge_sv1} shows that the trained neural network results are similar to MCMC for the US data from 1964Q2 to 2011Q1 (latest vintage in \cite{diebold2017real}). As for the previous model, Table \ref{tab::time_NLL} shows the NLL and MSE on 1,000,000 randomly generated datasets for future comparisons.

\subsection{Seasonal adjustment with structural breaks in seasonality}\label{subsec::3_3sa}
In addition to problems based on well-verified formulas for transition and observation equations, SBI is also suitable for those models where simulations are the primary focus. Usually, this occurs in models where deriving equations is too cumbersome or simply impossible to compute. Special cases are tasks where it is easy to generate a lot of different data which shows the model how it should behave in various situations.

As an example, we show how a quarterly seasonal adjustment model that considers structural shifts in the seasonal component can be built. As will be shown below, a traditional X13 ARIMA-SEATS procedure (see \cite{x13arimaseats}) does a poor job of this. An additional advantage is the automatically generated credible intervals.

\ref{app::D} presents the procedure for generating artificial series with a length of 40 to 80 quarters. In fact, it consists of generating a seasonal and non-seasonal component with the probability of a shift appearing in the seasonal part. Thus, the resulting series may not contain a break.

We compare NPE not with a sampling algorithm, but with X13 for measuring the quality, in contrast to the previous two models. The purpose of this experiment is to demonstrate how one can easily generate examples of model behavior, thus specifying an implicit Bayesian model. In practice, it is usually difficult to construct a fast MCMC algorithm in such cases. However, comparison with other algorithms that solve the same practical problem is of interest from the point of view of estimating the performance of the proposed algorithm.

\begin{figure}[H]
    \centering
    \includegraphics[width=\linewidth]{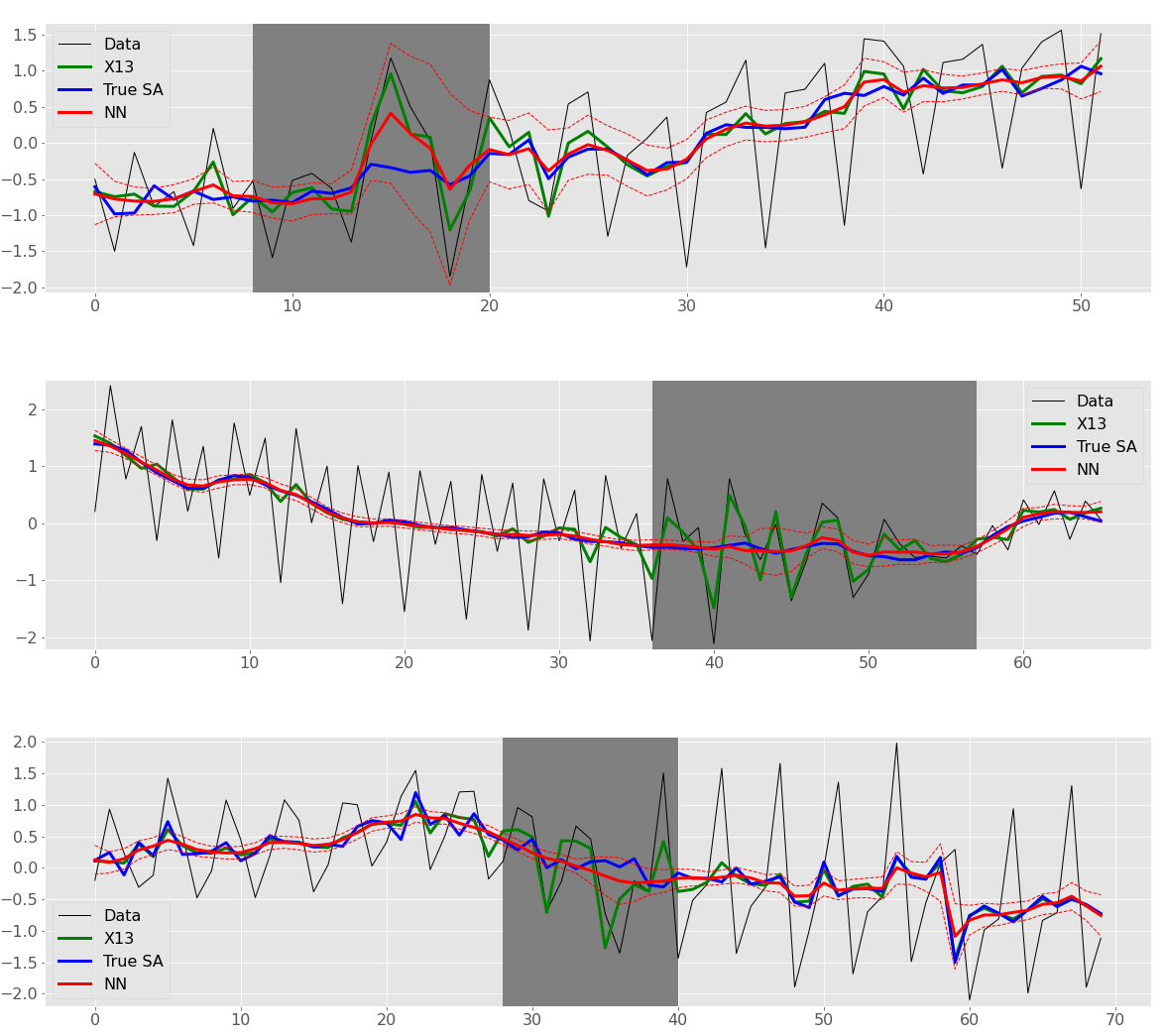}
    \caption{Comparison of NPE and X-13 ARIMA-SEATS on artificial series with a break in seasonality (mean $\pm$ 2 std)}
    \label{fig::npe_x13_comp}
\end{figure}

Figure \ref{fig::npe_x13_comp} shows random examples illustrating how the proposed procedure and X13 behave on series with a shift in seasonality (the gray area shows 1.5 years around the shift). X13 does not adequately cope with the task of detecting seasonality around the quarter of shift, while NPE, on the contrary, is robust. We calculated the MSE on 100,000 randomly generated runs for X13 and on 1,000,000 runs for the neural network. The error for NPE is smaller, as can be seen from Table \ref{tab::sa_x13_compare}. The smaller errors in comparison to X13 are not unexpected by themselves since cross-entropy optimization should provide the estimator with the lowest MSE on this dataset. However, the gap between the errors on the series with and without shifts further emphasizes that the algorithm proposed in the paper is more accurate than the alternative widely used among macroeconomists.

\begin{table}[H]
\centering
    \begin{tabular}{c| c c c}
     &	{Full sample}	& {With shifts}&	{Without shifts} \\
     \hline
     {NPE}	&$1.3\times10^{-3}$ &	$3.0\times10^{-3}$ &	$1.1\times10^{-3}$ \\
     {X13 ARIMA-SEATS} &	$5.4\times10^{-3}$ &	$25.0\times10^{-3}$ &	$3.1\times10^{-3}$ \\ 
     \hline
     \end{tabular}
    \caption{MSE for NPE and X13-ARIMA-SEATS on artificial data}\label{tab::sa_x13_compare}
\end{table}

\subsection{Computation and implementation time}\label{subsec::3_4comp}
As noted above, NPE works almost instantly due to possessing an amortization property. Depending on the task, calculating an approximation of the posterior distribution on the CPU (Intel(R) Core(TM) i7-8750H CPU @ 2.20GHz, 16GB RAM) takes tenths of a second for a pretrained model. Estimation of neural network parameters on Pytorch\footnote{Generation of artificial data for the model is run on the CPU and training on the GPU.}  (see \cite{NEURIPS2019_9015}) using GPU (NVIDIA GeForce RTX 2070) takes about 12, 12 and 2 hours for the stochastic volatility model, DSGE model and seasonal adjustment model, respectively. Comparing the running time for amortized algorithms with alternatives that do not possess such properties comes with difficulties. On the one hand, Bayesian model estimation for a fixed dataset takes less time in our examples when MCMC or VB algorithms are used (see Table \ref{tab::time_post_comp})\footnote{We choose the number of iterations based on convergence of parameters for MCMC (half of iterations is a burn-in period) and convergence of loss for VB.}. This is because the amortized algorithm requires pretraining of a neural network. On the other hand, a neural network trained once can be used for various datasets (including data of different lengths) which is an advantage in the case of multiple estimations.

\begin{table}[H]
\centering
    \begin{tabular}{c| c c | c c | c | c}
    &  \multicolumn{2}{c|}{ {NPE}}  & \multicolumn{2}{c|}{ {VB}} &  {MCMC} &  {Other} \\
    & {CPU} & {GPU} & {CPU} & {GPU} & {CPU} & {CPU} \\
    \hline
    {SV}&	0.42s&	0.08s&	1h27m&	16m	&19m	&---\\
    {SV-DSGE}&	0.14s&	0.02s&	--- &	--- &	9h18m &	--- \\
    {SA} &	0.14s&	0.05s&	--- &	--- &	--- &	0.27s \\
    \hline
     \end{tabular}
    \caption{Time of posterior computation per one dataset}\label{tab::time_post_comp}
\end{table}
The implementation of the NPE algorithm uses already available neural network libraries (coding a neural network architecture usually takes only tens of minutes since the layers of the neural network are already implemented in the respective libraries) and almost always turns out to be much easier than the implementation of MCMC and even sometimes variational estimation. The implementation of MCMC requires the derivation of the sampling algorithm and the writing of program code, which is often much more complicated than for NPE. Like NPE, stochastic VB algorithms are just as easy to implement in most cases (if automatic differentiation packages are used and there are no modules that require differentiation coding). The key difference is that the joint density of parameters, hidden states and data is used instead of data sampling procedure.

An important point for stochastic optimization algorithms (NPE and VB) is also the ability to almost effortlessly transfer the computation to the GPU.

\section{Related work}\label{sec::4rela_wrk}
The algorithm proposed in this paper is closely related to several directions presented in the literature. Our work is part of the literature on SBI (or likelihood-free inference) algorithms (see \cite{cranmer2020frontier}). Until recently, Bayesian direction in this field has developed mainly as approximate Bayesian computations (hereinafter, ABC). ABC approximate the likelihood functions by introducing an auxiliary likelihood function that depends on the distance between summary statistics of the data and the corresponding statistics of the simulated data. Classic sampling algorithms are then used (see \cite{sisson2018handbook}). The progress of machine learning algorithms, and in particular neural networks, has given rise to a whole family of algorithms that directly train posterior distributions (see \cite{papamakarios2016fast},  \cite{lueckmann2017flexible}, \cite{greenberg2019automatic}, \cite{durkan2020contrastive}), likelihood functions (see \cite{wood_statistical_2010}, \cite{lueckmann2018likelihood}, \cite{brehmer2020mining}, \cite{papamakarios2019sequential}) or likelihood function ratios (see \cite{brehmer2020mining}, \cite{hermans2020likelihood}, \cite{durkan2020contrastive}). These methods are not sensitive to tolerance hyperparameter and chosen distance between simulation data and summary statistics, contrary to ABC. However, to the best of our knowledge, apart from a few papers on probabilistic programming (see \cite{le2017inference}, \cite{baydin2019etalumis}, \cite{munk2019deep}), researchers concentrate mainly on parameters, not states. Research on probabilistic programming has two key differences from the approach proposed here. First, it uses the pretrained neural network as a proposal distribution for importance sampling, rather than directly to approximate the posterior distribution. Second, a neural network in probabilistic programming takes into account the relationship between variables to achieve smaller variance for the importance sampling weights, which is a considerably more difficult task in terms of optimization\footnote{It is also worth noting that probabilistic programming uses state sampling which depends on the sampled states of the previous period. This can lead to accumulation of approximation errors over long periods. Such an architecture is not quite suitable for direct approximation. However, this is not critical for subsequent resampling, especially if sequential importance sampling is used instead of importance sampling.}. Moreover, unlike this paper, implementation from scratch or modification of probabilistic programming algorithms for tasks that do not fit into the framework of standard libraries\footnote{See, for instance, PyProb.} is quite complicated, since it requires a deep knowledge of the addressing of random variables.

Our research is also closely related to the estimation of economic models through simulations. The simulated method of moments and its modifications (see \cite{mcfadden1989method}, \cite{10.2307/2951768}, \cite{gallant1996moments}) are common in the frequentist estimation of the structural parameters of models\footnote{See a list of applications in \cite{carrasco2002simulation}.}. There is a similar field of research that estimates model parameters based on minimizing various divergences between simulated and real data (see \cite{nickl2010efficient}, \cite{kaji2020adversarial}). \cite{gallant2009determination}\footnote{ \cite{gallant2013generalized} also developed a version of the Bayesian simulation method of moments based on a particle filter for models with hidden states.}  proposed a Bayesian version of the simulated method of moments. A recent paper by \cite{fen2022fast} uses sequential (non-amortized) NPE for Bayesian parameter estimation. As for SBI, not many papers devoted to simulated estimation of states rather than parameters exist. The closest known to us is the paper by \cite{gatti2020rising}, where the authors estimate states (output and investment gap forecasts) on artificial data using nonparametric kernel estimation. The idea is very close to SBI and to what is proposed here, but it is computationally difficult with a large number of hidden states as the authors themselves note.

Meta-learning (see \cite{finn2019meta}) is close to SBI in its mathematical formulation. Like SBI, meta-learning is based on the idea of learning from many similar tasks (see \cite{vinyals2016matching}). The key differences are purpose and data. Unlike SBI, meta-learning focuses on the task of predicting rather than finding the posterior distribution. Furthermore, meta-learning usually works with real data, not simulated ones.

Many works on variational autoencoders have been devoted to the amortization of finding the distribution of hidden states (see \cite{kingma2019introduction}). However, there are a number of differences from this paper. First, the data generation process is usually specified with a rather flexible model such as a neural network (see \cite{kingma2013auto}) or a Gaussian process (see \cite{dai2015variational}) rather than more classical models where the hidden states have greater identifiability and interpretability. In addition, the model is usually non-Bayesian in nature\footnote{Basically, neural networks are used as a data generation model, which by their nature are frequentist. Moreover, despite the fact that many applications use dropout for regularization (\cite{srivastava2014dropout}), which has a Bayesian interpretation (see \cite{kingma2015variational} and \cite{gal2016dropout}), the parameters are common to all data. This is ideologically different from the idea of amortizing models.}. Second, the loss function that is minimized is the KL divergence between the approximate posterior and posterior distributions, while in SBI it is the KL divergence between the posterior and approximate posterior distributions. The asymmetry of KL divergence leads to the fact that, with few exceptions (see \cite{tran2017hierarchical}), there are not enough simulations to train variational autoencoders and one must calculate the probabilities of the data generation process. Also, in the case of diagonal approximation (as in Section \ref{subsec::2_3marglik}), this leads to underestimation of variance (see \cite{blei2017variational}). Third, real, not artificial, data are used for training variational autoencoders as for meta-learning.

\section{Discussion}\label{sec::5disc}
As has been shown in many papers (see \cite{lueckmann2021benchmarking}), amortization leads to the need for longer training of SBI algorithms than their sequential counterparts. Despite this, we use the amortized NPE algorithm for two reasons. First, sequential SBI algorithms are usually applied to the problems of small dimension (the output of neural network dimension), and their adaptation to the high-dimensional problem of finding the posterior distribution of states is not trivial and requires solving more practical issues. In particular, if one tries to focus on marginal densities, as is done in this paper, the states generated for new rounds will not look like posterior distribution due to the lack of dependence between variables. The states will be quite noisy, which worsen convergence in most cases. Secondly, in contrast to some research on SBI, we do not set the task to finding the best algorithm under the constrained budget for the number of simulations (see \cite{lueckmann2021benchmarking}). Sequential algorithms give a significant gain for such tasks. However, the main goal of this paper is to build an algorithm that replaces long-running alternatives (as in the first two examples) or helps to estimate models where other algorithms fail (as in the third example)\footnote{It is implicitly assumed that a large number of model simulations can be performed in adequate time.} . Amortization is a great property that helps to solve this problem if the model is frequently re-estimated.

Many issues related to the estimation of the posterior distribution of states are beyond the scope of this paper and require further research. Some of them are discussed below.

The posterior distributions estimated using the proposed algorithm, although close to the MCMC results, nevertheless differ slightly. The results are slightly noisy when estimating stochastic volatility, while in the DSGE model they are biased. This signals an opportunity for further improvements of the neural networks by increasing the flexibility of the neural network architecture, the number of simulations or by modifying the training procedure. It is well known that for a certain learning rate schedule, stochastic optimization procedures converge to one of the local optima in the asymptotics (see Chapter 5 in \cite{kushner2003stochastic}). The exact (even local) optimum is not achieved with a finite number of iterations\footnote{This usually means that the learning rate does not tend to zero.}\footnote{ \cite{mandt2017stochastic} provide intuitions about the behavior of the estimation procedure at non-zero learning rates.}. An insufficiently flexible network and/or a small number of observations in the neighborhood of real data can lead to a situation where the model is unable to predict the posterior distribution accurately, even at the optimum\footnote{A good example of such type an improvement in the field of text analysis is the GPT-3 model (see \cite{brown}). It has reached a fundamentally new level compared to previous models due to an order of magnitude more parameters than previously used and a huge dataset.}. Moreover, the quality of the posterior distribution approximation is likely to deteriorate with increasing problem dimensionality. Hence, one of the main tasks for the future is to study the relationship between scalability, approximation quality and neural network training time.

The mean-field Gaussian approximation considered here is usually not a problem from a practical point of view, because in most cases, researchers are interested in the first and second moments of the marginal distributions of states. Although extensions have clear theoretical solutions (M-estimators for estimating other characteristics and more flexible families of distributions), their practical implementation requires further research\footnote{In a number of preliminary experiments that were not included in the paper, we saw that the quantile loss also shows good results for the marginal distributions.}.

We have bypassed the issues of forecasting and missing variables, which are related in the sense that the forecasting problem can be thought of as a problem of constructing a posterior distribution for the missing variables on the forecasting horizon. To deal with missing variables, models can be extended by introducing additional dummy variables as one of the inputs of the neural network, showing the presence of a miss, and/or by filling in the miss with learnable parameters as done in \cite{lueckmann2017flexible}.  A similar method or alternatives based on meta-learning ideas (see \cite{harrison2018meta}), where only the predicted variables are used as the neural network output, can be applied to build prediction models.

NPE has both advantages and disadvantages in terms of speed as shown in Section \ref{subsec::3_4comp}. Therefore, the choice to use NPE or not should depend on the situation. We recommend using the NPE algorithm if frequent re-estimation of the model is expected, or if alternative algorithms are slow, or fail to do the job at all. At the same time, we also advise to verify the trained algorithm before use by comparing it with alternative ones for approximating the posterior distribution (when they are not too slow). If such verification is impossible, it is recommended to carry out at least a visual analysis on artificially generated data.

\section{Conclusions}\label{sec::6concl}
The amortized simulation-based algorithm proposed in this paper for estimating hidden states of Bayesian state space models provides an alternative to already existing algorithms in this field. In contrast to many previous papers, we consider a new approach that approximates posterior marginal distribution of states and that does not rely on probability density functions for prior distributions, transition and observation equations, but that uses only simulations of artificial data.

The NPE algorithm shows results similar to other algorithms for the stochastic volatility and DSGE models but after training, it works nearly instantly. In addition, as shown in the example with seasonal adjustment, it also performs well on tasks where the Bayesian model is not specified directly but rather through the process of simulating various situations and correct behavior in them.

\newpage
\appendix
\renewcommand{\thesection}{Appendix \Alph{section}}
\renewcommand{\thesubsection}{\Alph{section}.\arabic{subsection}}
\renewcommand\thefigure{\Alph{section}1}
\renewcommand{\thealgo}{\Alph{section}.\arabic{algo}}

\section{Informal proofs}\label{app:A}
\subsection{NPE asymptotic }\label{app:A1}
Given a flexible parametric family $q_{\varphi} (\theta | y )$ and an infinite number of datasets, problem (1) becomes equivalent to the following problem:
\begin{multline}
    q^{*} = \argmin_q \mathbb{E}_{p(\theta, y)} \left( - log\, q(\theta|y) - log\, p(y) \right) = \\= 
    \argmin_q \int \int \bigl( - log\,q(\theta|y)  - log\,p(y)\bigr) p(\theta, y) d\theta dy  =\\ =
    \argmin_q \int \left( - \int log\,q(\theta|y)  p(\theta|y) d\theta \right) p(y)  dy
\end{multline}
Note that the optimization problem is split into a set of separate cross-entropy minimizations for each individual dataset -- $\int log\,q(\theta|y) \,p(\theta|y) d\theta$. The cross-entropy minimum is reached when the distributions coincide, which means that
$$
q^{*} = p(\theta | y)
$$
\subsection{NPE for states }\label{app:A2}
The only thing we need to prove for the validity of Algorithm \ref{algo::sbl_ss} is that the joint distribution of sampled states and data is the marginal distribution of the data generation process. We can then use the results of Appendix \ref{app:A1} by redefining $s$ as $\theta$.

The sample $\theta_i, s_i, y_i$ from Algorithm \ref{algo::sbl_ss} is identical to the sample from the data generation process, i.e. $\theta_i, s_i, y_i  \sim p(\theta,s,y)$. Integrating over $\theta$ we obtain that $s_i,y_i  \sim p(s,y)$, where  $p(s,y)$ is the marginal distribution of the data generation process.

\subsection{NPE for marginal distribution loss}\label{app:A3}
Consider an M-estimator loss function $m(x,s,y)$, where $y$ is the set of observed variables, $s$ is the set of hidden states, and $x$ is the set of posterior distribution characteristics estimated by the M-estimator. Let’s assume that $f_{\varphi} (y)$ is a parametric family of functions that maps dataset to the characteristics of the posterior distribution. With a flexible function $f$ and the number of simulations tending to infinity, we obtain 
$$
f^{*} = \argmin_{f} \mathbb{E}_{p(s,y)} \, m\left( f(y), s, y \right)  =
\argmin_{f} \int \left( \int m(f(y), s, y) p(s|y) ds\right) p(y) dy
$$
The problem splits into a set of minimizations for individual datasets $\int m(f(y),s,y)p(s|y)ds$ and $f^{*} (y)$ coincides with the M-estimator asymptotic value for each $y$ similarly to Appendix \ref{app:A1}. Thus, a set of classical M-estimators can be used to estimate the characteristics of the posterior distribution.

Means and standard deviations converge to their true values when an independent normal distribution is used as an approximation for the posterior distribution. So, if the function $f$ is sufficiently flexible and the number of simulations tends to infinity, the mean and standard deviations converge to their true values.
\newpage
\section{Stochastic volatility model}\label{app::B}
\subsection{Model}\label{app:B1}
\textit{Prior:}
$$
    \alpha \sim \mathcal{N}\left( 0, \sqrt{10}\right), \,\,\kappa \sim \mathcal{N}\left(0, \sqrt{10} \right), \,\, \psi \sim \mathcal{N}\left(0, \sqrt{10} \right) 
    $$
    $$
    \sigma = log\left( 1 + e^{\alpha} \right) , \,\, \rho = \frac{1}{1 + e^{-\psi}}
$$

\noindent\textit{Transition equation:}
\begin{align*}
&& && && 
SV_t \sim \mathcal{N} \left( \frac{\kappa}{2} (1 - \rho) + \rho SV_{t-1}, \frac{\sigma}{2} \right),  &&  t = 2, \ldots, T 
\end{align*}
$$
SV_1 \sim \mathcal{N} \left( \frac{\kappa}{2}, \frac{\sigma}{2\sqrt{1-\rho^2}} \right)
$$
\textit{Observation equation:}
$$
y_t \sim \mathcal{N}\left( 0, e^{SV_t}\right)
$$
\subsection{Architecture and learning algorithm}\label{app:B2}

\begin{figure}[H]
    \centering
    \includegraphics{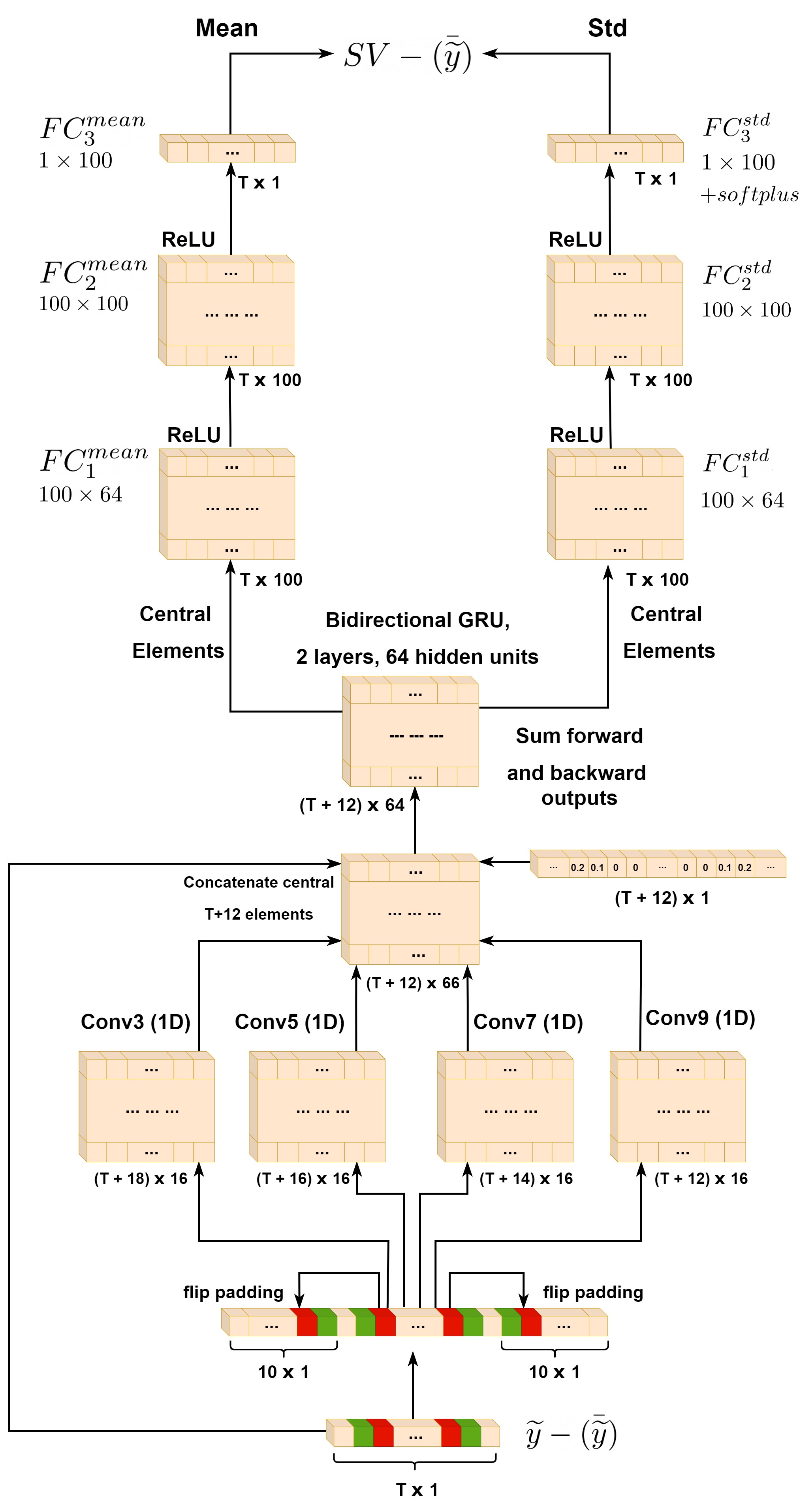}
    \caption{Neural network architecture for calculating mean and standard deviation in the stochastic volatility model}
    \label{fig::diag1}
\end{figure}

\begin{algo}{\textbf{Pretraining stochastic volatility model }}\label{algo::pretr_ss}
\vspace{0,25ex}\hrule\leavevmode \\
\normalfont  
$\left(B = 100,\,\, N_{sim} = 200,000,\,\, T_{lb} = 800,\,\, T_{ub} = 1,200, \,\,c = 10^{-30} \right)$\\
\\
For  $i = 1, \ldots, N_{sim}$:
\begin{enumerate}
    \item Draw $T_n$ from uniform discrete distribution $T \sim U(T_{lb}, T_{ub})$.
    \item Simulate $n^{th}$ batch:\\
    For  $b = 1, \ldots, B$:
    \begin{enumerate}
        {\setlength\itemindent{30pt}\item Draw $\sigma^b, \,\, \kappa^b, \,\, \rho^b$ from prior.}
        {\setlength\itemindent{30pt}\item Draw $SV^b = \left\{SV_{1}^{b}, \ldots, SV_{T_n}^{b} \right\}$ conditioned on $\sigma^b, \,\, \kappa^b, \,\, \rho^b$.}
        {\setlength\itemindent{30pt} \item Draw $\widetilde{y}^b = \left\{ log(c + |y_{1}^{b}|), \ldots, log(c + |y^{b}_{T_n}|) \right\}$ conditioned on $SV^b$.}
    \end{enumerate}
    $SV_{n}^{batch} = \left\{ \{SV^{1}, \widetilde{y}^{1} \}, \ldots, \{SV^{B}, \widetilde{y}^{B} \} \right\}$.
    \item Compute per state loss using architecture illustrated in Figure \ref{fig::diag1}:
    \begin{equation*}
        L_n = - \frac{1}{BT_n} \sum_{b=1}^{B}\sum_{t=1}^{T_n}log \,p \left(SV_{t}^{b} \bigg| m_t(\widetilde{y}^{b}, \varphi), \sigma_t(\widetilde{y}^{b}, \varphi)  \right) 
    \end{equation*}
    \item 	Make an optimization step with respect to $\varphi$ using ADAM algorithm.
\end{enumerate}
\end{algo}

ADAM (see \cite{kingma2014adam}) is applied with standard settings except for the learning rate:
\begin{equation*}
    \varepsilon_{n} = 
    \begin{cases}
        10^{-3}, & if \,\,\,\, n <3 \times 10^4\\
        10^{-4}, & if \,\,\,\, 3 \times 10^4 \leq n < 10^5\\
        10^{-5}, & if \,\,\,\, n \geq 10^5
    \end{cases}
\end{equation*}
\subsection{Alternative algorithm for stochastic volatility model}\label{app:B3}
The algorithms used for comparison are the adaptive MCMC and VB algorithms. The MCMC algorithm is based on an approximation of the logarithm of the square of a Gaussian random variable as a mixture of 7 normal distributions (see \cite{kim1998stochastic}). For this purpose, the observation equation is rewritten as:
$$
log\, y_{t}^{2} = 2SV_t + \sum_{k=1}^{7} z_t e_t^k 
$$
$$
e^k \sim \mathcal{N} (\mu_k, \sigma_k)
$$
$$
p(z_t = k) = \omega_k
$$
\noindent where $\mu_k$, $\sigma_k$, $\omega_k$ are constants defined in \cite{kim1998stochastic}. Algorithm \ref{algo::mcmc_sv} describes the complete procedure.

\begin{algo}{\textbf{Adaptive MCMC algorithm for stochastic volatility model }}\label{algo::mcmc_sv}
\vspace{0.25ex}\hrule\leavevmode \\
\normalfont  
$\left( N_{sim} = 2,000, \,\, c = 1.5, \,\, \Sigma_0 = 0.1\times I \right)$\\
\\
For $i = 1, \ldots, N_{sim}$:
\begin{enumerate}
    \item For $t=1,\ldots,T$ draw discrete approximation to chi-squared distribution:
    $$
    z_{t}^{n} \sim p\left(z_t | SV_{t}^{n-1}, y_t \right)
    $$
    \item 	Draw parameters $\theta = \{ \alpha, \kappa, \psi \}$ using Random Walk Metropolis-Hastings algorithm with adaptive proposal (see \cite{adapt}):
    $$
    q(\theta' | \theta_{n-1}) = 0.95 \times \mathcal{N} \left(\theta^{n-1}, \frac{c^2}{3}\Sigma_{n-1} \right) + 0.05\times  \mathcal{N} \left(\theta^{n-1}, \frac{0.01^2}{3} I \right)
    $$
    and acceptance rate:
    $$
    ar_n = \min\left( \frac{p(y_1, \ldots, y_T | z_{1}^{n}, \ldots, z_{T}^{n}, \theta') p(\theta')}{p(y_1, \ldots, y_T | z_{1}^{n}, \ldots, z_{T}^{n}, \theta^{n-1}) p(\theta^{n-1})}, 1 \right)
    $$
    \item Draw stochastic volatility:
    $$
    SV_{1}^{n}, \ldots, SV_{T}^{n} \sim p\left( SV_{1}^{}, \ldots, SV_{T}^{} \bigg| z_{1}^{n}, \ldots, z_{T}^{n}, y_1, \ldots, y_T, \theta^n  \right)
    $$
\end{enumerate}
\end{algo}
Note that after introducing the variables $z_{1}^{}, \ldots, z_{T}^{}$, steps 2 and 3 of Algorithm \ref{algo::mcmc_sv} can be implemented via standard Kalman filter and Kalman sampler procedures (see \cite{durbin2002simple}).

The VB estimator uses a Gaussian approximation algorithm where the precision matrix is sparse (see \cite{tan2018gaussian}). 20,000 iterations of the ADAM algorithm with a learning rate of 0.001 and a batch size of 100 are used for training.

\newpage
\newpage

\section{Stochastic volatility DSGE model}\label{app::C}
\subsection{Model}\label{app:C1}
The DSGE model, similar to \cite{diebold2017real} is considered, but with several modifications. First, we remove inflation expectations from the observed variables and exclude the shock of the inflation target from the model to avoid dealing with omitted variables. Second, prior distributions are slightly changed for processes associated with stochastic volatility to make the simulations more realistic.\\
\\
\textit{Prior:}
$$
\tau \sim \mathcal{N}(1.5, 0.36),\,\,
\nu_l \sim \mathcal{G}(2, 0.75),\,\,
\iota \sim \mathcal{B}(0.5, 0.15),\,\,
\zeta \sim \mathcal{B} (0.5, 0.1) ,
$$
$$
    \psi_1 \sim \mathcal{N}(1.5, 0.25) ,\,\,
    \psi_2 \sim \mathcal{N}(0.12, 0.05) ,\,\,
    -400\, log(\beta) \sim \mathcal{G}(1, 0.4) ,\,\,
    400 \,log(\pi_{*}) \sim \mathcal{G}(2.48, 0.4),
    $$
    $$
    100 \,log(\gamma) \sim \mathcal{N}(0.4,0.1),\,\,  
    \rho_R \sim \mathcal{B}(0.5, 0.2),\,\,
     \rho_g \sim \mathcal{B} (0.5, 0.2),\,\, 
     \varphi_z \sim U(-1, 1),
     $$
     $$
     (100 \,\sigma_R)^2 \sim \mathcal{IG} (0.1, 2),\,\,
     (10 \, \sigma_g)^2 \sim \mathcal{IG} (0.1, 2),\,\,
     (10\,  \sigma_z)^2 \sim \mathcal{IG} (0.1, 2),
     $$
     $$
     (0.2\,\sigma_{R}^{SV})^2 \sim \mathcal{IG}(0.05, 2),\,\,
          (0.2\, \sigma_{g}^{SV})^2 \sim \mathcal{IG}(0.05, 2),\,\, 
     (0.2\,\sigma_{z}^{SV})^2 \sim \mathcal{IG}(0.05, 2),
     $$
     $$
     \rho_{R}^{SV} \sim \mathcal{N}(0.9, 0.07),\,\,
     \rho_{g}^{SV} \sim \mathcal{N}(0.9, 0.07),\,\,
     \rho_{z}^{SV} \sim \mathcal{N}(0.9, 0.07)
$$

\noindent where $\mathcal{N}(a,b), \mathcal{B}(a,b), \mathcal{G}(a,b)$ are normal, beta and gamma distributions with mean $a$ and standard deviation $b$, $U(a,b)$ is a uniform distribution with upper and lower bounds $a$ and $b$, $\mathcal{IG}(a,b)$ is an inverse gamma distribution with probability density $p(x) \sim x^{-b-1} e^{-\frac{ba^2}{2x}}$.
\\
\\
\noindent\textit{Transition equations:}
\\
The transition equations are given in the form:
\setlength{\abovedisplayskip}{0pt} \setlength{\abovedisplayshortskip}{0pt}

$$
s_t \sim \mathcal{N} \left(  A(\theta) \, s_{t-1},\,\, B(\theta)\, diag(e^{SV_t}) \, B^{T} (\theta) \right), \,\,\,\,\,\,  t = 2, \ldots, T
$$
$$
SV_t = \left\{ SV_{t}^{g}, SV_{t}^{z}, SV_{t}^{R}\right\}
$$
$$
s_1 \sim \mathcal{N} \left( 0, \,\, P(\theta) \right)
$$
$$
\,\,SV_{t}^{i} \sim \mathcal{N} \left( \rho_{i}^{SV} SV_{t-1}^{i}, \,\, \sigma_{i}^{SV} \right),  \,\,\,\,\,\,  t = 2, \ldots, T,\,\, i \in \{ g, z, R\}
$$
$$
SV_{t}^{i} \sim \mathcal{N} \left( 0, \,\, \frac{\sigma_{i}^{SV}}{\sqrt{1 - \left( \rho_{i}^{SV} \right)^2 }}  \right), \,\,\,\,\,\,  i \in \{ g, z, R\}
$$
where $\theta = \left\{ \tau, \nu_l, \iota, \zeta, \psi_1, \psi_2, \beta, \pi_{*}, \gamma, \rho_{R}, \rho_{g}, \varphi_{z}, \sigma_{R}, \sigma_{g}, \sigma_{z} \right\}$, 
$s_t = \left\{ y_t, c_t, g_t, \pi_t, R_t, z_t, dy_t \right\}$, $P(\theta)$ is the solution of equation: 

$$
P(\theta) = A(\theta) \, P(\theta) \, A^{T}(\theta) + B(\theta) \, B^{T}(\theta) 
$$
and $A(\theta)$ and $B(\theta)$ is a stable solution\footnote{Parameters for which there are many stable solutions or there is no stable solution are excluded.}  of the following linear system of stochastic discrete equations:
$$
    c_t = \mathbb{E}_t \left( c_{t+1} + z_{t+1} \right) - \frac{1}{\tau} \left( R_t - \mathbb{E}_t \pi_{t+1} \right) 
    $$
    $$
    \pi_t = \frac{\iota}{1 + \beta \iota} \pi_{t-1} + \frac{\beta}{1 + \beta \iota} \mathbb{E}_t \pi_{t+1} + \frac{(1 - \zeta \beta)(1 - \zeta)}{(1 + \beta \iota)\zeta} \left( c_t + \nu_l y_t \right)
    $$
    $$
    y_t = c_t + g_t
    $$
    $$
    R_t = \rho_R R_{t-1} + (1- \rho_R) \left(\psi_1 \pi_t + \psi_2  (y_t - y_{t-1} + z_t ) \right) + \sigma_R e^{R}_{t} $$
    $$
    z_t = -\varphi_z z_{t-1} + \sigma_z e_{t}^{z} 
    $$
    $$
    g_t = \rho_g g_{t-1} + \sigma_g e^{g}_{t} 
    $$
    $$
    dy_t = y_t - y_{t-1} + z_{t} 
    $$
\noindent where $y_t,\,c_t,\,g_t,\,\pi_t,\,R_t,\,z_t,\,dy_t$ are variables that correspond to deviations from the steady state of output, consumption, exogenous process responsible for the share of government consumption, inflation, interest rate, exogenous technological process and GDP growth, $e_{t}^{R},\,e_{t}^{z},\,e_{t}^{g}$ are monetary policy, technology and government consumption shocks.
\\

\noindent\textit{Observation equations:}

The observation equations have the form:

$$
obs_t = 
\begin{bmatrix}
dy_{t}^{obs} \\
\pi_{t}^{obs} \\
R_{t}^{obs}
\end{bmatrix}
=
\begin{bmatrix}
100 \,log\,\gamma \\
100 \,log\, \pi_{*} \\
100 \,(log\,\gamma + log \, \pi_{*} - log\, \beta)
\end{bmatrix}
+ 
\begin{bmatrix}
100 \,dy_t\\
100 \,\pi_{t} \\
100 \,R_{t}
\end{bmatrix}
$$
where $dy_{t}^{obs}$ is quarterly real GDP growth, $\pi_{t}^{obs}$ is quarterly price growth and $R_{t}^{obs}$ is interest rate in quarterly terms.

\subsection{Architecture and learning algorithm}\label{app:C2}
\begin{figure}[H]
    \centering
    \includegraphics{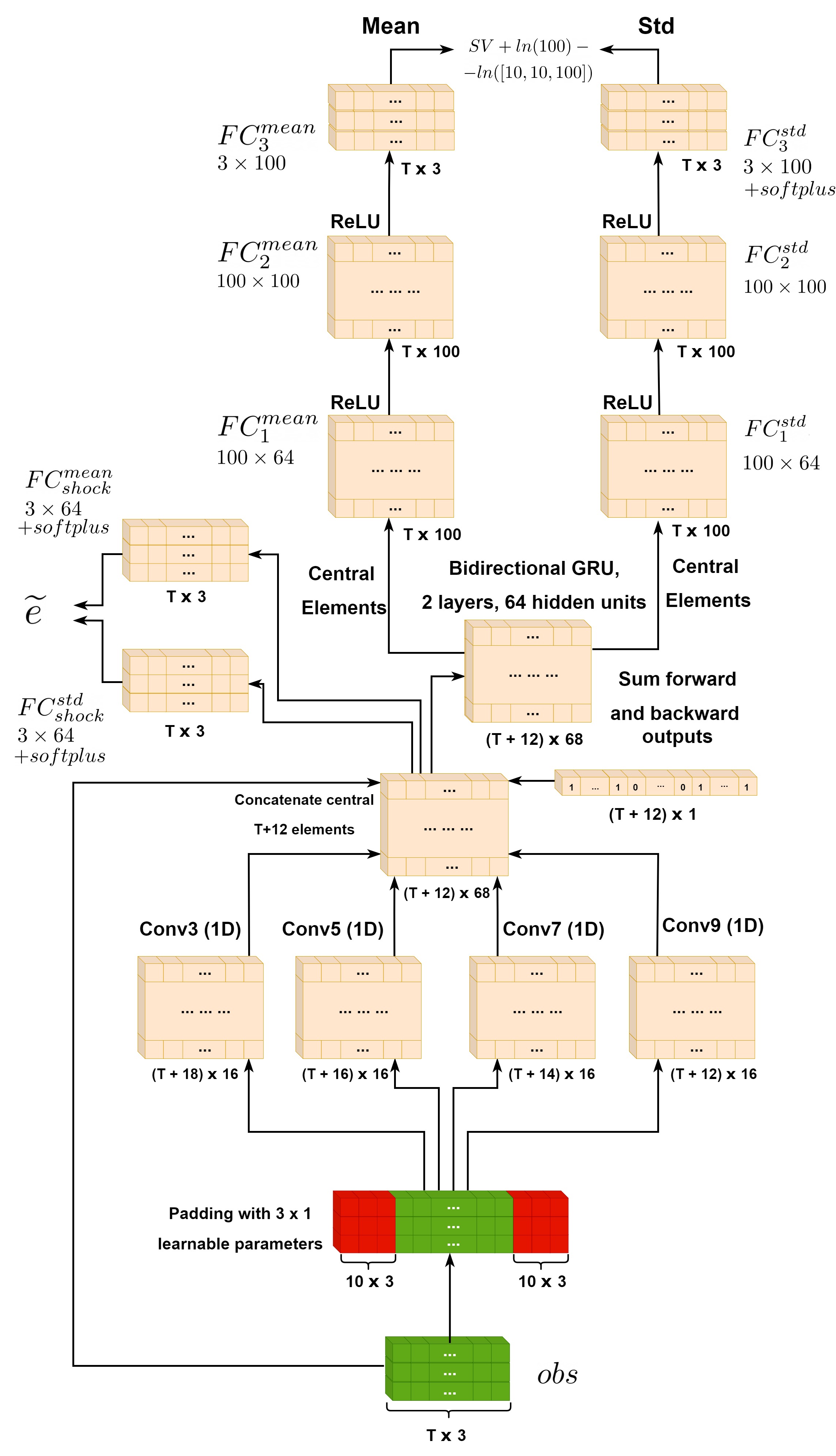}
    \caption{Neural network architecture for calculating mean and standard deviation in the SV-DSGE model}
    \label{fig::diag2}
\end{figure}

\begin{algo}{\textbf{Pretraining stochastic volatility DSGE model  }}\label{algo::sbi_dsge}
\vspace{0.25ex}\hrule\leavevmode \\
\normalfont  
$\left( N_{presim} = 1,000,000, \,\,B = 100, \,\, N_{sim} = 500, 000, \,\,
T_{lb} = 180, \,\,T_{ub} = 200,\,\, w = 1,\,\, c = 10^{-30} \right)$\\
\\
\noindent Set $n_{presim} = 0, \theta = \{\}, A = \{\},B = \{\}.$\\
\noindent While  $n_{presim} < N_{presim}$:
\begin{enumerate}
    \item Draw $\theta_{n_{presim}}$ from prior.
    \item Solve DSGE\footnote{\cite{anderson1985linear} algorithm is applied.}.
    \item If solution is stable\footnote{In our case, there is almost no non-unique stable or unstable solutions. See \cite{lueckmann2017flexible} as a one of examples how to deal with situations where certain regions of the parameter space are implausible.} 	append $\theta_{n_{presim}},\,A_{n_{presim}},\, B_{n_{presim} }$ in $\theta,\,A,\,B$ and increment $n_{presim}$ by $1$.
\end{enumerate}
For  $n = 1, \ldots, N_{sim}$:
\begin{enumerate}
    \item	Draw $T_n$ from uniform discrete distribution $T \sim U(T_{lb},T_{ub} )$.
    \item 	Simulate $n^{th}$ batch:\\
    For \,\, $b = 1, \ldots, B$:
    \begin{enumerate}
        {\setlength\itemindent{30pt}\item Draw $\{ \theta^{b},\, A^{b}, \, B^{b}\}$ uniformly from $\{ \theta^{},\, A^{}, \, B^{}\}$ and $\sigma_{g}^{SV,b}, \, \sigma_{z}^{SV,b}, \, \sigma_{R}^{SV,b},$\\ $ \rho_{g}^{SV,b}, \rho_{z}^{SV,b}, \, \rho_{R}^{SV, b}$ from prior. }
        {\setlength\itemindent{30pt}\item Draw $SV^b = \{ SV_{1}^{b}, \ldots, SV_{T_n}^{b}\}, s^b = \{ s_{1}^{b}, \ldots, s_{T_n}^{b}\}$ and $\widetilde{e}^{b} = \bigg\{\big\{ log(c+|e^{R,b}_{1}|), log(c+|e^{z,b}_{1}|), log(c+|e^{g,b}_{1}|)\big\}, \ldots, \big\{ log(c+|e^{R,b}_{T_n}|), log(c+|e^{z,b}_{T_n}|), log(c+|e^{g,b}_{T_n}|)\big\}  \bigg\}$ conditioned on draw from 2a.}
        {\setlength\itemindent{30pt}\item Draw $obs^b = \{ obs^{b}_{1}, \ldots,  obs^{b}_{T_n}\}$ conditioned on $SV^b$, $s^b$ and $\theta^b$.}
    \end{enumerate}
$SV_{n}^{batch} = \bigg\{ \{ SV^{1}, obs^{1} \}, \ldots, \{ SV^{B}, obs^{B} \} \bigg\}$.
\item Compute loss using architecture illustrated in Figure \ref{fig::diag2}:
    \begin{multline*}
        L_n = - \frac{1}{3\, BT_n} \sum_{b=1}^{B}\sum_{t=1}^{T_n}\sum_{i \in \{ g,z,R \}} 
         \bigg(
        log\, p\bigg(SV_{t}^{i, b}  \bigg| m_{t,i}^{SV}(obs^b, \varphi),  \sigma_{t,i}^{SV}(obs^b, \varphi) \bigg)  + \\  w\,log \, p\bigg( \widetilde{e}_{t}^{i,b}  \bigg| m_{t,i}^{e}(obs^b, \varphi),  \sigma_{t,i}^{e}(obs^b, \varphi) \bigg)   \bigg)
    \end{multline*}
\item Make an optimization step with respect to $\varphi$ using the ADAM algorithm.
\end{enumerate}

\end{algo}
The learning rate, $\varepsilon_{n}$, for the ADAM algorithm has the following schedule:
\begin{equation*}
    \varepsilon_{n} = 
    \begin{cases}
        10^{-3}, & if \,\,\,\, n <3 \times 10^4\\
        10^{-4}, & if \,\,\,\, 3 \times 10^4 \leq n < 10^5\\
        10^{-5}, & if \,\,\,\, 10^5 \leq n < 2 \times 10^5\\
        10^{-6}, & if \,\,\,\, 2 \times 10^5 \leq n < 3.5 \times 10^5\\
        3 \times 10^{-6}, & if \,\,\,\, n \geq 3.5 \times 10^5\\
    \end{cases}
\end{equation*}

\subsection{Alternative algorithm for stochastic volatility DSGE model}\label{app:C3}
Algorithm \ref{algo::sbi_dsge} is compared with an adaptive MCMC algorithm similar to that proposed by \cite{justiniano2008time} and \cite{diebold2017real}. The key difference is the replacement of Random Walk Metropolis-Hastings step for sampling parameters of stochastic volatilities (with marginalized states) by Gibbs sampling step.

\begin{algo}{\textbf{Adaptive MCMC algorithm for stochastic volatility DSGE model   }}\label{algo::mcmc_dsge}
\vspace{0.25ex}\hrule\leavevmode \\
\normalfont 
$\left( N_{sim} = 100,000, \,\, c = 1.5, \,\,
\Sigma_{0}^{\theta} = 0.1 \times I, \,\, \Sigma_{0}^{\theta^{SV}} = 0.1 \times I \right)$\\
\\
For  $n = 1, \ldots, N_{sim}$:
\begin{enumerate}
    \item		Draw parameters $\theta$ using Random Walk Metropolis-Hastings algorithm with adaptive proposal:
    $$
    q(\theta' | \theta_{n-1}) = 0.95 \times \mathcal{N} \left(\theta^{n-1}, \frac{c^2}{15}\Sigma_{n-1}^{\theta} \right) + 0.05\times  \mathcal{N} \left(\theta^{n-1}, \frac{0.01^2}{15} I \right)
    $$
    and acceptance rate:
    $$
    ar_n = \min\left( \frac{p(obs_1, \ldots, obs_T | SV_{1}^{n-1}, \ldots, SV_{T}^{n-1}, \theta') p(\theta')}{p(obs_1, \ldots, obs_T | SV_{1}^{n-1}, \ldots, SV_{T}^{n-1}, \theta^{n-1}) p(\theta^{n-1})}, 1 \right)
    $$
    \item Draw errors $e_{1}, \ldots, e_{T}$:
    $$
    e_{1}^{n}, \ldots, e_{T}^{n} \sim p\bigg( e_1, \ldots, e_T \bigg|  obs_1, \ldots, obs_T, SV_{1}^{n-1}, \ldots, SV_{T}^{n-1}, \theta^{n} \bigg)
    $$
    \item For  $t = 1, \ldots, T$:
    $$
    z_{t}^{i,n} \sim p \left( z_{t}^{i} \bigg| SV_{1}^{n-1}, \ldots, SV_{T}^{n-1}, e_{1}^{n}, \ldots, e_{T}^{n} \right), \,\,\, i \in \{g,z,R\}
    $$
    \item 	Draw parameters $\theta^{SV}=\bigg\{ \sigma_{g}^{SV},\sigma_{z}^{SV}, \sigma_{R}^{SV}, \rho_{g}^{SV},\rho_{z}^{SV},\rho_{R}^{SV} \bigg\}$ using Random Walk Metropolis-Hastings algorithm with adaptive proposal:
    
    $$
    q( {\theta'}^{SV} | \theta_{n-1}^{SV}) = 0.95 \times \mathcal{N} \left(\theta^{SV, n-1}, \frac{c^2}{6}\Sigma_{n-1}^{\theta^{SV}} \right) + 0.05\times  \mathcal{N} \left(\theta^{SV,n-1}, \frac{0.01^2}{6} I \right)
    $$
    and acceptance rate:
    $$
    ar_n = \min\left( \frac{p(e_{1}^{n}, \ldots, e_{T}^{n} | z_{1}^{n}, \ldots, z_{T}^{n}, {\theta'}^{SV}) p( {\theta'}^{SV} )}{p(e_{1}^{n}, \ldots, e_{T}^{n} | z_{1}^{n}, \ldots, z_{T}^{n}, \theta^{SV, n-1}) p( \theta^{SV,n-1} )}, 1 \right)
    $$
    \item	Draw stochastic volatility:
    $$
    SV_{1}^{n}, \ldots, SV_{T}^{n} \sim p\left( SV_{1}, \ldots, SV_{T} \bigg| z_{1}^{n}, \ldots, z_{T}^{n}, e_{1}, \ldots, e_{T}, \theta^{SV, n} \right)
    $$
\end{enumerate}
\end{algo}

\newpage

\section{ {Seasonal adjustment with structural breaks in \\ \,\,\,\,\, seasonality}}\label{app::D}
The data generation process is not directly specified here, unlike in previous models. Instead, we describe the data generation procedure:

\begin{algo}{\textbf{Data generator with breaks in seasonality    }}\label{algo::sa_dgp}
\vspace{0.25ex}\hrule\leavevmode \\
\normalfont  
$\left( T_{lb} = 40,\,\, T_{ub} = 80, \,\, T^{period} = 4,\,\,B=100 \right)$\\
\begin{enumerate}
    \item 	Draw $T_n$ from uniform discrete distribution $T \sim U(T_{lb},T_{ub} )$.
    \item Simulate batch components:\\
    {For} $b = 1,\ldots, B$:
    \begin{enumerate}
        {\setlength\itemindent{30pt}\item Generate non-seasonal component $NS^b$:
        $$
        e_{t}^{b} \sim Student\left( 0,1,3+ |\eta_t| \right), \,\,\,\,\,\, t = -199,\ldots, T_n
        $$
        $$
        \eta_t \sim \mathcal{N}(0, 3),\,\,\,\,\,\,  t = -199,\ldots, T_n
        $$
        $$
        e_{t}^{shift, b} \sim \mathcal{N} (0, 20) \times Bernouli(0.01), \,\,\,\,\,\,  t = -196,\ldots, T_n
        $$
       $$
        \rho_{AR,1}^{b}, \rho_{AR,2}^{b}, \rho_{AR,3}^{b}, \rho_{MA,1}^{b}, \rho_{MA,2}^{b}, \rho_{MA,3}^{b} \sim Bernouli(0.5) \times U(-0.5, 0.98) 
        $$
        \begin{multline*}
         \varepsilon_{t} = e_{t}^{b} + (\rho_{MA,1}^{b}+\rho_{MA,2}^{b}+\rho_{MA,3}^{b}) e_{t-1}^{b} + \\
         (\rho_{MA,1}^{b}\rho_{MA,2}^{b}+\rho_{MA,2}^{b}\rho_{MA,3}^{b}+\rho_{MA,1}^{b}\rho_{MA,3}^{b}) e_{t-2}^{b}  \\
         + \rho_{MA,1}^{b}\rho_{MA,2}^{b}\rho_{MA,3}^{b}e_{t-3}^{b} + e_{t}^{shift,b}, \,\,\,\,\,\,\,\,\,\,\,\,\,\,\,\,\,\,\,\,\,\,\,\,t = -196,\ldots,T_n\\
         x_{t}^{b} = (\rho_{AR,1}^{b}+\rho_{AR,2}^{b}+\rho_{AR,3}^{b}) x_{t-1}^{b} - (\rho_{AR,1}^{b}\rho_{AR,2}^{b}+\rho_{AR,2}^{b}\rho_{AR,3}^{b}+\rho_{AR,1}^{b}\rho_{AR,3}^{b}) x_{t-2}^{b}  \\
         + \rho_{AR,1}^{b}\rho_{AR,2}^{b}\rho_{AR,3}^{b}x_{t-3}^{b} + \varepsilon_t, \,\,\,\,\,\,\,\,\,\,\,\,\,\,\,\,\,\,\,\,\,\,\,\,\,\,\,t = -196,\ldots,T_n
        \end{multline*}
        $$
         x_{-199}^{b}, x_{-198}^{b}, x_{-197}^{b} = 0
        $$
        $$
        c^{b} \sim \mathcal{N} (0, 0.005), \,\,\,\,\,\,\,\,\,\, scale^{b} \sim \mathcal{N} \left(0, \frac{0.007}{std(x^b)} \right)
        $$
        $$
        NS^{*b} = \{ c^b + scale^b x_{-199}, \ldots, c^{b} + scale^{b} x_{T_n} \}
        $$
        $$
        I^{integrated,b} \sim Bernouli(0.5)
        $$
        $$
        NS^b = I^{integrated,b}\times cumsum(NS^{*b}) + (1 - I^{integrated,b}) NS^{*b}
        $$}
    {\setlength\itemindent{30pt}\item Generate seasonal component $S^b$: \\
    $$
    \sigma^b \sim \mathcal{N} \left(0, \frac{0.2}{\sqrt{40}} \right)
    $$
    $$
        I^{shift,b}_{t} \sim Bernouli(0.01), \,\,\,\, \,\,\,\,  t = -199,\ldots,T_n
        $$
        $$
        z_{t}^{b} = I^{shift,b}_{t-3}I^{shift,b}_{t-2}I^{shift,b}_{t-1}I^{shift,b}_{t}, \,\,\,\, \,\,\,\,  t = -196,\ldots,T_n
        $$
        $$
        e_{t}^{S,b} \sim \mathcal{N} (0,1), \,\,\,\, \,\,\,\,  t = -196,\ldots,T_n 
        $$
        $$
        s_{-199}^{b}, s_{-198}^{b},  s_{-197}^{b} \sim \mathcal{N}(0,1)
        $$
        $$
        s_{t}^{b} = - (1-z_{t}^{b}) (s_{t-1}^{b} + s_{t-2}^{b} + s_{t-3}^{b} + \sigma^b e^{S,b}_{t}) + z_{t}^{b}e^{S,b}_{t}
        $$
        $$
        scale^{S,b} \sim \mathcal{N} \left(0, 3\,std(NS^{*b}) \right)
        $$
        $$
        S^b = \{ scale^{S,b} s^{b}_{-199}, \ldots, scale^{S,b} s^{b}_{T_n} \}
   $$
   }
\end{enumerate}
    \item Create batch of size $2B$:\\
    {For} \,\, $b = 1,\ldots, B$:
    $$
    y^{*b} = S^{b} + NS^{b}
    $$
    $$
    y^b = \bigg\{ \frac{y_{1}^{b} - mean(y^{*b})}{ std(y^{*b})} , \ldots, \frac{y_{T}^{b} - mean(y^{*b})}{ std(y^{*b})} \bigg\}
    $$
    $$
    y^{B+b} = \bigg\{ \frac{y_{T}^{b} - mean(y^{*b})}{ std(y^{*b})} , \ldots, \frac{y_{1}^{b} - mean(y^{*b})}{ std(y^{*b})} \bigg\}
    $$
    $$
    sa^{b} = \bigg\{ \frac{NS_{1}^{b} - mean(y^{*b})}{ std(y^{*b})} , \ldots, \frac{NS_{T}^{b} - mean(y^{*b})}{ std(y^{*b})} \bigg\}
    $$
    $$
    sa^{B+b} = \bigg\{ \frac{NS_{T}^{b} - mean(y^{*b})}{ std(y^{*b})} , \ldots, \frac{NS_{1}^{b} - mean(y^{*b})}{ std(y^{*b})} \bigg\}
    $$
\item[] $
    y = \{y^{1}, \ldots, y^{2B} \}, \,\,\,\,\,\,\,\,\,\,  sa =\{ sa^{1},\ldots,sa^{2B}\}.
$
\end{enumerate}

\end{algo}

The neural network estimation algorithm approximating the mean and standard deviation of the posterior distribution is similar to those described for other models.

\begin{figure}[H]
    \centering
    \includegraphics{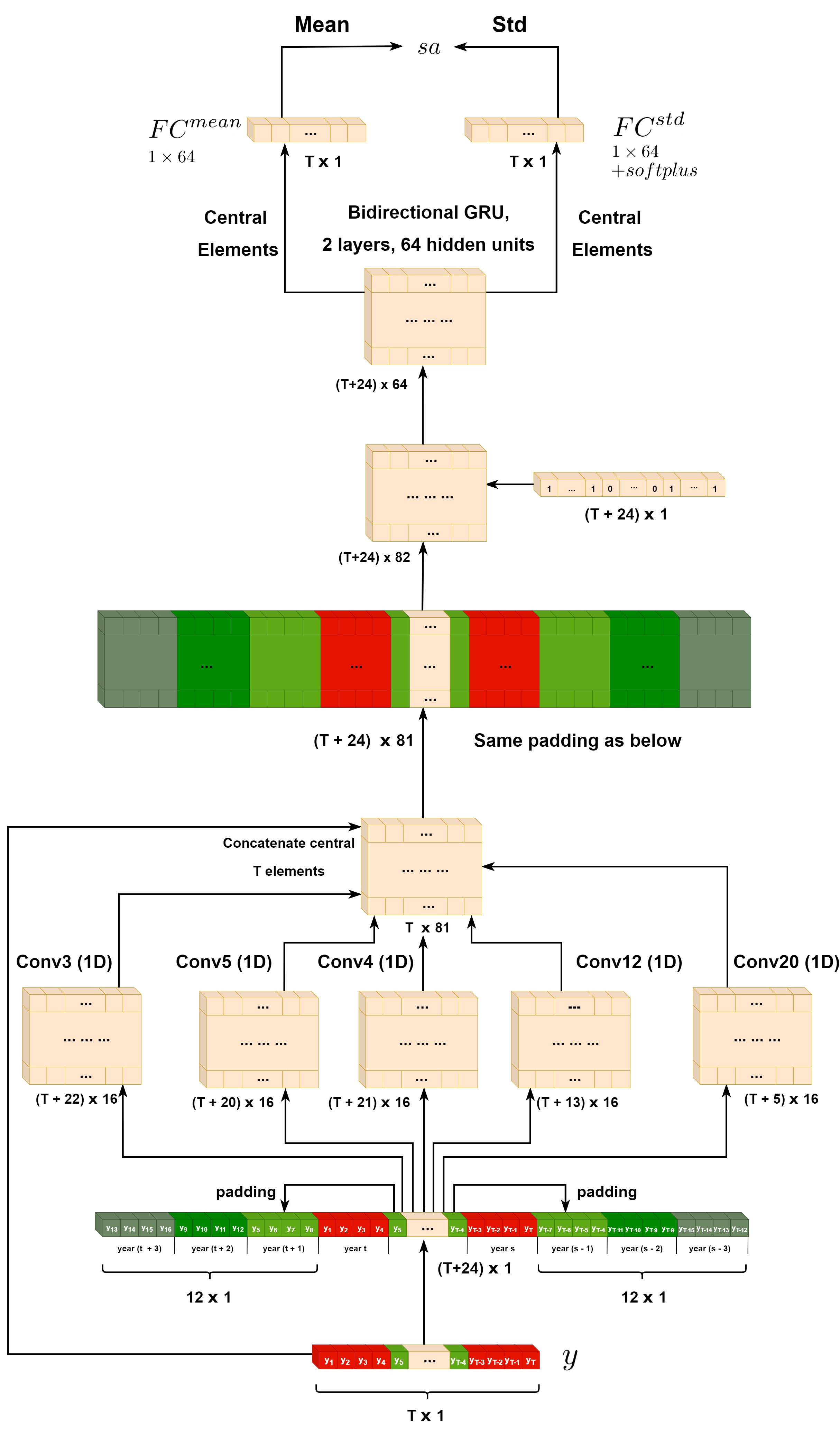}
    \caption{Neural network architecture for calculating mean and standard deviation in the SA model}
    \label{fig::diag3}
\end{figure}

\begin{algo}{\textbf{Pretraining seasonal adjustment with structural breaks in seasonality  }}\label{algo::sbi_sa}
\vspace{0.25ex}\hrule\leavevmode \\
\normalfont  
$\left( N_{sim} = 100,000\right)$\\
\\
{For}  $i = 1, \ldots, N_{sim}$:
\begin{enumerate}
    \item Draw $T_n$ from uniform discrete distribution $T \sim U(T_{lb}, T_{ub})$.
    \item Simulate $n^{th}$ batch using Algorithm \ref{algo::sa_dgp}:\\
    $$sa_{n}^{batch} = \left\{ \{sa^{1}, {y}^{1} \}, \ldots, \{sa^{2B}, {y}^{2B} \} \right\}$$
    \item Compute per state loss using architecture illustrated in Figure \ref{fig::diag3}:
    \begin{equation*}
        L_n = - \frac{1}{2BT_n} \sum_{b=1}^{2B}\sum_{t=1}^{T_n}log \,p\left( sa_{t}^{b} \bigg| m_t({y}^{b}, \varphi), \sigma_t({y}^{b}, \varphi)  \right) 
    \end{equation*}
    \item 	Make an optimization step with respect to $\varphi$ using the ADAM algorithm.
\end{enumerate}
\end{algo}

The schedule for the ADAM algorithm is defined as:
\begin{equation*}
    \varepsilon_{n} = 
    \begin{cases}
        10^{-3}, & if \,\,\,\, n <1.5 \times 10^4\\
        10^{-4}, & if \,\,\,\, 1.5 \times 10^4 \leq n < 5 \times 10^4\\
        10^{-5}, & if \,\,\,\, n \geq 5 \times 10^4
    \end{cases}
\end{equation*}

\newpage
\newpage


\newpage
\begin{thebibliography}{}

\bibitem[Anderson and Moore, 1985]{anderson1985linear}
Anderson, G. and Moore, G. (1985).
\newblock {A} {L}inear {A}lgebraic {P}rocedure for {S}olving {L}inear {P}erfect
  {F}oresight {M}odels.
\newblock {\em Economics Letters}, 17(3):247--252.

\bibitem[Andrieu et~al., 2010]{andrieu2010particle}
Andrieu, C., Doucet, A., and Holenstein, R. (2010).
\newblock {P}article {M}arkov {C}hain {M}onte {C}arlo {M}ethods.
\newblock {\em Journal of the Royal Statistical Society: Series B (Statistical
  Methodology)}, 72(3):269--342.

\bibitem[Baydin et~al., 2019]{baydin2019etalumis}
Baydin, A., Shao, L., Bhimji, W., Heinrich, L., Meadows, L., Liu, J., Munk, A.,
  Naderiparizi, S., Gram-Hansen, B., Louppe, G., Ma, M., Zhao, X., Torr, P.,
  Lee, V., Cranmer, K., Prabhat, and Wood, F. (2019).
\newblock Etalumis: Bringing probabilistic programming to scientific simulators
  at scale.
\newblock {\em In {P}roceedings of the {I}nternational {C}onference for {H}igh
  {P}erformance {C}omputing, {N}etworking,{S}torage and {A}nalysis (SC19)}.

\bibitem[Beal, 2003]{beal2003variational}
Beal, M.~J. (2003).
\newblock Variational {A}lgorithms for {A}pproximate {B}ayesian {I}nference.
\newblock {\em PhD Thesis, Gatsby Computational Neuroscience Unit, University
  College London}.

\bibitem[Beaumont et~al., 2002]{beaumont2002approximate}
Beaumont, M.~A., Zhang, W., and Balding, D.~J. (2002).
\newblock {A}pproximate {B}ayesian {C}omputation in {P}opulation {G}enetics.
\newblock {\em Genetics}, 162(4):2025--2035.

\bibitem[Blanchard and Kahn, 1980]{blanchard1980solution}
Blanchard, O.~J. and Kahn, C.~M. (1980).
\newblock The {S}olution of {L}inear {D}ifference {M}odels under {R}ational
  {E}xpectations.
\newblock {\em Econometrica}, 48(5):1305--1311.

\bibitem[Blei et~al., 2017]{blei2017variational}
Blei, D.~M., Kucukelbir, A., and McAuliffe, J.~D. (2017).
\newblock Variational {I}nference: {A} {R}eview for {S}tatisticians.
\newblock {\em Journal of the American Statistical Association},
  112(518):859--877.

\bibitem[Blum and Fran{\c{c}}ois, 2010]{blum2010non}
Blum, M.~G. and Fran{\c{c}}ois, O. (2010).
\newblock Non-{L}inear {R}egression {M}odels for {A}pproximate {B}ayesian
  {C}omputation.
\newblock {\em Statistics and Computing}, 20(1):63--73.

\bibitem[Brehmer et~al., 2020]{brehmer2020mining}
Brehmer, J., Louppe, G., Pavez, J., and Cranmer, K. (2020).
\newblock Mining {G}old from {I}mplicit {M}odels to {I}mprove {L}ikelihood-free
  {I}nference.
\newblock {\em Proceedings of the National Academy of Sciences},
  117(10):5242--5249.

\bibitem[Brown et~al., 2020]{brown}
Brown, T., Mann, B., Ryder, N., Subbiah, M., Kaplan, J., Dhariwal, P.,
  Neelakantan, A., Shyam, P., Sastry, G., Askell, A., Agarwal, S.,
  Herbert-Voss, A., Krueger, G., Henighan, T., Child, R., Ramesh, A., Ziegler,
  D., Wu, J., Winter, C., Hesse, C., Chen, M., Sigler, E., Litwin, M., Gray,
  S., Chess, B., Clark, J., Berner, C., McCandlish, S., Radford, A., Sutskever,
  I., and Amodei, D. (2020).
\newblock {Language Models are Few-Shot Learners}.
\newblock {\em arXiv:2005.14165v4.}

\bibitem[Carrasco and Florens, 2002]{carrasco2002simulation}
Carrasco, M. and Florens, J.-P. (2002).
\newblock Simulation-{B}ased {M}ethod of {M}oments and {E}fficiency.
\newblock {\em Journal of Business \& Economic Statistics}, 20(4):482--492.

\bibitem[Carriero et~al., 2019]{carriero2019large}
Carriero, A., Clark, T.~E., and Marcellino, M. (2019).
\newblock Large {B}ayesian {V}ector {A}utoregressions with {S}tochastic
  {V}olatility and {N}on-{C}onjugate {P}riors.
\newblock {\em Journal of Econometrics}, 212(1):137--154.

\bibitem[Casella and George, 1992]{casella1992explaining}
Casella, G. and George, E.~I. (1992).
\newblock Explaining the {G}ibbs {S}ampler.
\newblock {\em The American Statistician}, 46(3):167--174.

\bibitem[Chib and Greenberg, 1995]{chib1995understanding}
Chib, S. and Greenberg, E. (1995).
\newblock Understanding the {M}etropolis-{H}astings {A}lgorithm.
\newblock {\em The American Statistician}, 49(4):327--335.

\bibitem[Chiu et~al., 2012]{chiu2012estimating}
Chiu, C.~W., Eraker, B., Foerster, A.~T., Kim, T.~B., and Seoane, H.~D. (2012).
\newblock Estimating {VAR}’s {S}ampled at {M}ixed or {Irregular Spaced
  Frequencies: A Bayesian Approach.}
\newblock {\em Federal Reserve Bank of Kansas City RWP}, 11-11.

\bibitem[Chopin et~al., 2013]{chopin2013smc2}
Chopin, N., Jacob, P.~E., and Papaspiliopoulos, O. (2013).
\newblock {SMC\^\,2: an Efficient Algorithm for Sequential Analysis of State
  Space Models}.
\newblock {\em Journal of the Royal Statistical Society: Series B (Statistical
  Methodology)}, 75(3):397--426.

\bibitem[Cranmer et~al., 2020]{cranmer2020frontier}
Cranmer, K., Brehmer, J., and Louppe, G. (2020).
\newblock {The Frontier of Simulation-Based Inference}.
\newblock {\em Proceedings of the National Academy of Sciences},
  117(48):30055--30062.

\bibitem[Dai et~al., 2016]{dai2015variational}
Dai, Z., Damianou, A., Gonz{\'a}lez, J., and Lawrence, N. (2016).
\newblock {Variational Auto-encoded Deep Gaussian Processes}.
\newblock {\em International Conference on Learning Representations}.

\bibitem[Del~Moral et~al., 2006]{del2006sequential}
Del~Moral, P., Doucet, A., and Jasra, A. (2006).
\newblock {Sequential Monte Carlo Samplers}.
\newblock {\em Journal of the Royal Statistical Society: Series B (Statistical
  Methodology)}, 68(3):411--436.

\bibitem[Deli~Gatti and Grazzini, 2020]{gatti2020rising}
Deli~Gatti, D. and Grazzini, J. (2020).
\newblock {Rising to the Challenge: Bayesian Estimation and Forecasting
  Techniques for Macroeconomic Agent Based Models}.
\newblock {\em Journal of Economic Behavior \& Organization}, 178:875--902.

\bibitem[Diebold et~al., 2017]{diebold2017real}
Diebold, F.~X., Schorfheide, F., and Shin, M. (2017).
\newblock {Real-time Forecast Evaluation of DSGE Models with Stochastic
  Volatility}.
\newblock {\em Journal of Econometrics}, 201(2):322--332.

\bibitem[Duffie and Singleton, 1993]{10.2307/2951768}
Duffie, D. and Singleton, K.~J. (1993).
\newblock {Simulated Moments Estimation of Markov Models of Asset Prices}.
\newblock {\em Econometrica}, 61(4):929--952.

\bibitem[Durbin and Koopman, 2002]{durbin2002simple}
Durbin, J. and Koopman, S.~J. (2002).
\newblock {A Simple and Efficient Simulation Smoother for State Space Time
  Series Analysis}.
\newblock {\em Biometrika}, 89(3):603--616.

\bibitem[Durkan et~al., 2020]{durkan2020contrastive}
Durkan, C., Murray, I., and Papamakarios, G. (2020).
\newblock {On Contrastive Learning for Likelihood-Free Inference}.
\newblock {\em In Proceedings of the 36th International Conference on Machine
  Learning.}

\bibitem[Fen, 2022]{fen2022fast}
Fen, C. (2022).
\newblock {Fast Simulation-Based Bayesian Estimation of Heterogeneous and
  Representative Agent Models using Normalizing Flow Neural Networks}.
\newblock {\em arXiv preprint arXiv:2203.06537}.

\bibitem[Fernández-Villaverde et~al., 2016]{FERNANDEZVILLAVERDE2016527}
Fernández-Villaverde, J., Rubio-Ramírez, J., and Schorfheide, F. (2016).
\newblock {Solution and Estimation Methods for DSGE Models}.
\newblock {\em Handbook of Macroeconomics}, 2:527--724.

\bibitem[Finn and Levine, 2019]{finn2019meta}
Finn, C. and Levine, S. (2019).
\newblock {Meta-learning: from Few-Shot Learning to Rapid Reinforcement
  Learning}.
\newblock {\em The International Conference on Machine Learning (ICML)},
  Tutorial.

\bibitem[Gal and Ghahramani, 2016]{gal2016dropout}
Gal, Y. and Ghahramani, Z. (2016).
\newblock {Dropout as a Bayesian Approximation: Representing Model Uncertainty
  in Deep Learning}.
\newblock {\em Neural Information Processing Systems}.

\bibitem[Gallant et~al., 2013]{gallant2013generalized}
Gallant, A.~R., Giacomini, R., and Ragusa, G. (2013).
\newblock {Generalized Method of Moments with Latent Variables}.
\newblock {\em CEPR Discussion Paper}, DP9692.

\bibitem[Gallant and McCulloch, 2009]{gallant2009determination}
Gallant, A.~R. and McCulloch, R.~E. (2009).
\newblock {On the Determination of General Scientific Models with Application
  to Asset Pricing}.
\newblock {\em Journal of the American Statistical Association},
  104(485):117--131.

\bibitem[Gallant and Tauchen, 1996]{gallant1996moments}
Gallant, A.~R. and Tauchen, G. (1996).
\newblock {Which Moments to Match?}
\newblock {\em Econometric theory}, 12(4):657--681.

\bibitem[Greenberg et~al., 2019]{greenberg2019automatic}
Greenberg, D., Nonnenmacher, M., and Macke, J. (2019).
\newblock {Automatic Posterior Transformation for Likelihood-Free Inference}.
\newblock {\em In Proceedings of the 36th International Conference on Machine
  Learning}.

\bibitem[Gunawan et~al., 2021]{gunawan2021variational}
Gunawan, D., Kohn, R., and Nott, D. (2021).
\newblock {Variational Bayes Approximation of Factor Stochastic Volatility
  Models}.
\newblock {\em International Journal of Forecasting}, 37(4):1355--1375.

\bibitem[Hamilton, 1989]{10.2307/1912559}
Hamilton, J.~D. (1989).
\newblock {A New Approach to the Economic Analysis of Nonstationary Time Series
  and the Business Cycle}.
\newblock {\em Econometrica}, 57(2):357--384.

\bibitem[Harrison et~al., 2020]{harrison2018meta}
Harrison, J., Sharma, A., and Pavone, M. (2020).
\newblock {Meta-Learning Priors for Efficient Online Bayesian Regression}.
\newblock {\em International Workshop on the Algorithmic Foundations of
  Robotics XIII}, 48:318--337.

\bibitem[Hermans et~al., 2020]{hermans2020likelihood}
Hermans, J., Begy, V., and Louppe, G. (2020).
\newblock {Likelihood-Free MCMC with Amortized Approximate Ratio Estimators}.
\newblock {\em In Proceedings of the 37th International Conference on Machine
  Learning}.

\bibitem[Hodrick and Prescott, 1997]{hodrick1997postwar}
Hodrick, R.~J. and Prescott, E.~C. (1997).
\newblock {Postwar US Business Cycles: an Empirical Investigation}.
\newblock {\em Journal of Money, credit, and Banking}, 29(1):1--16.

\bibitem[Hoffman et~al., 2013]{hoffman2013stochastic}
Hoffman, M.~D., Blei, D.~M., Wang, C., and Paisley, J. (2013).
\newblock {Stochastic Variational Inference}.
\newblock {\em Journal of Machine Learning Research}, 14(1):1303--1347.

\bibitem[Justiniano and Primiceri, 2008]{justiniano2008time}
Justiniano, A. and Primiceri, G.~E. (2008).
\newblock {The Time-Varying Volatility of Macroeconomic Fluctuations}.
\newblock {\em American Economic Review}, 98(3):604--41.

\bibitem[Kaji et~al., 2020]{kaji2020adversarial}
Kaji, T., Manresa, E., and Pouliot, G. (2020).
\newblock {An Adversarial Approach to Structural Estimation}.
\newblock {\em arXiv preprint arXiv:2007.06169}.

\bibitem[Kim et~al., 1998]{kim1998stochastic}
Kim, S., Shephard, N., and Chib, S. (1998).
\newblock {Stochastic Volatility: Likelihood Inference and Comparison with ARCH
  Models}.
\newblock {\em The Review of Economic Studies}, 65(3):361--393.

\bibitem[Kingma and Ba, 2014]{kingma2014adam}
Kingma, D.~P. and Ba, J. (2014).
\newblock {Adam: A method for stochastic optimization}.
\newblock {\em International Conference on Learning Representations}.

\bibitem[Kingma et~al., 2015]{kingma2015variational}
Kingma, D.~P., Salimans, T., and Welling, M. (2015).
\newblock {Variational Dropout and the Local Reparameterization Trick}.
\newblock {\em In Advances in Neural Information Processing Systems},
  2575--2583.

\bibitem[Kingma and Welling, 2014]{kingma2013auto}
Kingma, D.~P. and Welling, M. (2014).
\newblock {Auto-Encoding Variational Bayes}.
\newblock {\em International Conference on Learning Representations}.

\bibitem[Kingma and Welling, 2019]{kingma2019introduction}
Kingma, D.~P. and Welling, M. (2019).
\newblock {An Introduction to Variational Autoencoders}.
\newblock {\em Foundations and Trends in Machine Learning}, 12(4):307--392.

\bibitem[Klein, 2000]{klein2000using}
Klein, P. (2000).
\newblock {Using the Generalized Schur Form to Solve a Multivariate Linear
  Rational Expectations Model}.
\newblock {\em Journal of Economic Dynamics and Control}, 24(10):1405--1423.

\bibitem[Koop and Korobilis, 2012]{korobilis}
Koop, G. and Korobilis, D. (2012).
\newblock {Forecasting Inflation Using Dynamic Model Averaging}.
\newblock {\em International Economic Review}, 53(3):867--886.

\bibitem[Kushner and Yin, 2003]{kushner2003stochastic}
Kushner, H. and Yin, G.~G. (2003).
\newblock {Stochastic Approximation and Recursive Algorithms and Applications}.
\newblock {\em Springer Science \& Business Media}, 35.

\bibitem[Laubach and Williams, 2003]{laubach2003}
Laubach, T. and Williams, J.~C. (2003).
\newblock {Measuring the Natural Rate of Interest}.
\newblock {\em Review of Economics and Statistics}, 85(4):1063--1070.

\bibitem[Le et~al., 2017]{le2017inference}
Le, T.~A., Baydin, A.~G., and Wood, F. (2017).
\newblock {Inference Compilation and Universal Probabilistic Programming}.
\newblock {\em In Proceedings of the 20th International Conference on
  Artificial Intelligence and Statistics (AISTATS)}.

\bibitem[Lind{\'e} et~al., 2016]{linde2016challenges}
Lind{\'e}, J., Smets, F., and Wouters, R. (2016).
\newblock {Challenges for Central Banks’ Macro Models}.
\newblock {\em Handbook of Macroeconomics}, 2:2185--2262.

\bibitem[Lueckmann et~al., 2018]{lueckmann2018likelihood}
Lueckmann, J.-M., Bassetto, G., Karaletsos, T., and Macke, J.~H. (2018).
\newblock {Likelihood-Free Inference with Emulator Networks}.
\newblock {\em In Proceedings of the 1st Symposium on Advances in Approximate
  Bayesian Inference}.

\bibitem[Lueckmann et~al., 2021]{lueckmann2021benchmarking}
Lueckmann, J.-M., Boelts, J., Greenberg, D., Goncalves, P., and Macke, J.
  (2021).
\newblock {Benchmarking Simulation-Based Inference}.
\newblock {\em Proceedings of the 24th International Conference on Artificial
  Intelligence and Statistics (AISTATS)}.

\bibitem[Lueckmann et~al., 2017]{lueckmann2017flexible}
Lueckmann, J.-M., Goncalves, P.~J., Bassetto, G., {\"O}cal, K., Nonnenmacher,
  M., and Macke, J.~H. (2017).
\newblock {Flexible Statistical Inference for Mechanistic Models of Neural
  Dynamics}.
\newblock {\em Advances in Neural Information Processing Systems},
  30:1289--1299.

\bibitem[Lux, 2018]{lux2018estimation}
Lux, T. (2018).
\newblock {Estimation of Agent-Based Models Using Sequential Monte Carlo
  Methods}.
\newblock {\em Journal of Economic Dynamics and Control}, 91:391--408.

\bibitem[Mandt et~al., 2017]{mandt2017stochastic}
Mandt, S., Hoffman, M.~D., and Blei, D.~M. (2017).
\newblock {Stochastic Gradient Descent as Approximate Bayesian Inference}.
\newblock {\em Journal of Machine Learning Research}, 18:1--35.

\bibitem[McFadden, 1989]{mcfadden1989method}
McFadden, D. (1989).
\newblock {A Method of Simulated Moments for Estimation of Discrete Response
  Models without Numerical Integration}.
\newblock {\em Econometrica}, 57(5):995--1026.

\bibitem[Munk et~al., 2022]{munk2019deep}
Munk, A., Zwartsenberg, B., Scibior, A., Baydin, A.~G., Stewart, A.~L.,
  Fernlund, G., Poursartip, A., and Wood, F. (2022).
\newblock {Probabilistic Surrogate Networks for Simulators with Unbounded
  Randomness}.
\newblock {\em arXiv preprint arXiv:1910.11950}.

\bibitem[Neal, 1996]{neal2012bayesian}
Neal, R.~M. (1996).
\newblock {Bayesian Learning for Neural Networks}.
\newblock {\em Springer-Verlag, Lecture Notes in Statistics}, №118.

\bibitem[Nickl and P{\"o}tscher, 2010]{nickl2010efficient}
Nickl, R. and P{\"o}tscher, B.~M. (2010).
\newblock {Efficient Simulation-Based Minimum Distance Estimation and Indirect
  Inference}.
\newblock {\em Mathematical Methods of Statistics}, 19(4):327--364.

\bibitem[Orphanides and Van~Norden, 2002]{orphanides2002unreliability}
Orphanides, A. and Van~Norden, S. (2002).
\newblock {The Unreliability of Output-Gap Estimates in Real Time}.
\newblock {\em Review of Economics and Statistics}, 84(4):569--583.

\bibitem[Otrok and Whiteman, 1998]{otrok1998bayesian}
Otrok, C. and Whiteman, C.~H. (1998).
\newblock {Bayesian Leading Indicators: Measuring and Predicting Economic
  Conditions in Iowa}.
\newblock {\em International Economic Review}, (4):997--1014.

\bibitem[Papamakarios and Murray, 2016]{papamakarios2016fast}
Papamakarios, G. and Murray, I. (2016).
\newblock {Fast $\varepsilon$-free Inference of Simulation Models with Bayesian
  Conditional Density Estimation}.
\newblock {\em In Advances in Neural Information Processing Systems},
  29:1028--1036.

\bibitem[Papamakarios et~al., 2019]{papamakarios2019sequential}
Papamakarios, G., Sterratt, D., and Murray, I. (2019).
\newblock {Sequential Neural Likelihood: Fast Likelihood-Free Inference with
  Autoregressive Flows}.
\newblock {\em In Proceedings of the 22nd International Conference on
  Artificial Intelligence and Statistics (AISTATS)}.

\bibitem[Paszke et~al., 2019]{NEURIPS2019_9015}
Paszke, A., Gross, S., Massa, F., Lerer, A., Bradbury, J., Chanan, G., Killeen,
  T., Lin, Z., Gimelshein, N., Antiga, L., Desmaison, A., Kopf, A., Yang, E.,
  DeVito, Z., Raison, M., Tejani, A., Chilamkurthy, S., Steiner, B., Fang, L.,
  Bai, J., and Chintala, S. (2019).
\newblock {PyTorch: An Imperative Style, High-Performance Deep Learning
  Library}.
\newblock {\em {In Advances in Neural Information Processing Systems}},
  32:8024--8035.

\bibitem[Primiceri, 2005]{primiceri2005time}
Primiceri, G.~E. (2005).
\newblock {Time Varying Structural Vector Autoregressions and Monetary Policy}.
\newblock {\em The Review of Economic Studies}, 72(3):821--852.

\bibitem[Rezende and Mohamed, 2015]{rezende2015variational}
Rezende, D. and Mohamed, S. (2015).
\newblock {Variational Inference with Normalizing Flows}.
\newblock {\em In Proceedings of the 32nd International conference on machine
  learning (ICML)}.

\bibitem[Roberts and Rosenthal, 2009]{adapt}
Roberts, G.~O. and Rosenthal, J.~S. (2009).
\newblock Examples of adaptive mcmc.
\newblock {\em Journal of Computational and Graphical Statistics},
  18(2):349--367.

\bibitem[Schorfheide and Song, 2015]{schorfheide2015}
Schorfheide, F. and Song, D. (2015).
\newblock {Real-Time Forecasting With a Mixed-Frequency VAR}.
\newblock {\em Journal of Business \& Economic Statistics}, 33(3):366--380.

\bibitem[Schorfheide and Song, 2021]{schorfheide2021real}
Schorfheide, F. and Song, D. (2021).
\newblock {Real-time Forecasting with a (Standard) Mixed-Frequency VAR during a
  Pandemic}.
\newblock {\em NBER Working Papers}, № 29535.

\bibitem[Sims, 2002]{sims2002solving}
Sims, C.~A. (2002).
\newblock {Solving Linear Rational Expectations Models}.
\newblock {\em Computational economics}, 20(1-2):1.

\bibitem[Sisson et~al., 2018]{sisson2018handbook}
Sisson, S.~A., Fan, Y., and Beaumont, M. (2018).
\newblock {Handbook of Approximate Bayesian Computation}.
\newblock {\em Chapman and Hall{/}CRC Press}.

\bibitem[Smets and Wouters, 2003]{smets2003estimated}
Smets, F. and Wouters, R. (2003).
\newblock {An Estimated Dynamic Stochastic General Equilibrium Model of the
  Euro Area}.
\newblock {\em Journal of the European Economic Association}, 1(5):1123--1175.

\bibitem[Smets and Wouters, 2007]{smets2007shocks}
Smets, F. and Wouters, R. (2007).
\newblock {Shocks and Frictions in US Business Cycles: A Bayesian DSGE
  Approach}.
\newblock {\em American Economic Review}, 97(3):586--606.

\bibitem[Srivastava et~al., 2014]{srivastava2014dropout}
Srivastava, N., Hinton, G., Krizhevsky, A., Sutskever, I., and Salakhutdinov,
  R. (2014).
\newblock {Dropout: a Simple Way to Prevent Neural Networks from Overfitting}.
\newblock {\em The Journal of Machine Learning Research}, 15(1):1929--1958.

\bibitem[Stock and Watson, 2011]{stock2011dynamic}
Stock, J.~H. and Watson, M. (2011).
\newblock {Dynamic Factor Models}.
\newblock {\em In Clements M.J. and D.F. Hendry Oxford Handbook on Economic
  Forecasting.}

\bibitem[Tan et~al., 2020]{tan2020conditionally}
Tan, L.~S., Bhaskaran, A., and Nott, D.~J. (2020).
\newblock {Conditionally Structured Variational Gaussian Approximation with
  Importance Weights}.
\newblock {\em Statistics and Computing}, 30(5):1255--1272.

\bibitem[Tan and Nott, 2018]{tan2018gaussian}
Tan, L.~S. and Nott, D.~J. (2018).
\newblock {Gaussian Variational Approximation with Sparse Precision Matrices}.
\newblock {\em Statistics and Computing}, 28(2):259--275.

\bibitem[Tran et~al., 2017]{tran2017hierarchical}
Tran, D., Ranganath, R., and Blei, D. (2017).
\newblock {Hierarchical Implicit Models and Likelihood-Free Variational
  Inference}.
\newblock {\em Advances in Neural Information Processing Systems}, 30.

\bibitem[{US Census Bureau}, 2017]{x13arimaseats}
{US Census Bureau} (2017).
\newblock {X-13 ARIMA-SEATS Reference Manual}.
\newblock {\em US Census Bureau}.

\bibitem[Van~der Vaart, 2000]{van2000asymptotic}
Van~der Vaart, A.~W. (2000).
\newblock {\em Asymptotic statistics}, volume~3.
\newblock Cambridge university press.

\bibitem[Vinyals et~al., 2016]{vinyals2016matching}
Vinyals, O., Blundell, C., Lillicrap, T., Wierstra, D., et~al. (2016).
\newblock {Matching Networks for One Shot Learning}.
\newblock {\em Advances in Neural Information Processing Systems}, 29.

\bibitem[Wainwright and Jordan, 2008]{wainwright}
Wainwright, M. and Jordan, M. (2008).
\newblock {Graphical Models, Exponential Families, and Variational Inference}.
\newblock {\em Foundations and Trends in Machine Learning}, 1(1-2):1--305.

\bibitem[Walsh, 2010]{walsh2010}
Walsh, C. (2010).
\newblock {Monetary Theory and Policy}.
\newblock {\em MIT Press}, 3.

\bibitem[Wood, 2010]{wood_statistical_2010}
Wood, S.~N. (2010).
\newblock {Statistical Inference for Noisy Nonlinear Ecological Dynamic
  Systems}.
\newblock {\em Nature}, 466(7310):1102--1104.

\end{thebibliography}
\end{document}